\documentclass[
reprint,
superscriptaddress,
amsmath,amssymb,
prb,
floatfix,
showkeys
]{revtex4-2}
\usepackage{empheq}
\usepackage{float}
\usepackage{lipsum}
\usepackage{mathtools, cuted}
\usepackage{esvect}
\usepackage{xcolor}
\usepackage[caption = false]{subfig}
\usepackage{graphicx}
\usepackage{dcolumn}
\usepackage{bm}
\usepackage{hyperref}
\usepackage{physics}
\usepackage[mathlines]{lineno}

\newcommand{\C}{\mathbb{C}} 
\newcommand{\R}{\mathbb{R}} 
\newcommand{\N}{\mathbb{N}} 
\newcommand{\Z}{\mathbb{Z}} 

\newcommand{\brac}[1]{\left( #1 \right)} 

\begin{document}

\preprint{APS/123-QED}

\title{Multifractal Properties of Tribonacci Chains}

\author{Julius Krebbekx}
\affiliation{Institute of Theoretical Physics, Utrecht University.}%
\affiliation{Mathematical Institute, Utrecht University.}%

\author{Anouar Moustaj}
 \affiliation{Institute of Theoretical Physics, Utrecht University.}%

 \author{Karma Dajani}
\affiliation{Mathematical Institute, Utrecht University.}%

\author{Cristiane Morais Smith}
\affiliation{Institute of Theoretical Physics, Utrecht University.}%

\date{\today}

\begin{abstract}
We introduce two 1D tight-binding models based on the Tribonacci substitution, the hopping and on-site Tribonacci chains, which generalize the Fibonacci chain. For both hopping and on-site models, a perturbative real-space renormalization procedure is developed. We show that the two models are equivalent at the fixed point of the renormalization group flow, and that the renormalization procedure naturally gives the Local Resonator Modes. Additionally, the Rauzy fractal, inherent to the Tribonacci substitution, is shown to serve as the analog of conumbering for the Tribonacci chain. The renormalization procedure is used to repeatedly subdivide the Rauzy fractal into copies of itself, which can be used to describe the eigenstates in terms of Local Resonator Modes. Finally, the multifractal dimensions of the energy spectrum and eigenstates of the hopping Tribonacci chain are computed, from which it can be concluded that the Tribonacci chains are critical.

\end{abstract}

\keywords{Aperiodic; Quasicrystal; Multifractal Spectrum; Tribonacci; Rauzy Fractal}

\maketitle


\section{\label{sec:Intro}Introduction 
}


The description of electrons in solids using Bloch’s theorem has allowed for a profound understanding of the electronic band structure of regular crystalline materials \cite{BLOCH1929}. The discovery of quasicrystals \cite{SHECHTMAN}, aperiodic structures that break translational symmetry, has pushed the field forward. The Penrose tilings \cite{BRUIJN} or the aperiodic mono-tile discovered recently by Smith et al. \cite{smith2023MONOTILE} are some of the typical examples that have fascinated physicists and mathematicians for years. Quasi-crystalline lattices have been also experimentally realized using different quantum-simulator platforms, such as ultracold atoms \cite{SCHNEIDER} or photonics~\cite{BLOCH_ZILBERBERG}. 

The advent of topological insulators has reiterated the importance of periodicity in solids because translation invariance is at the core of the topological classification of these materials \cite{TKNN, ATLAND_ZIRNBAUER, RYU2010}. It remains an open question how the notion of topology translates to aperiodic structures such as quasicrystals, where translation invariance is often replaced by scale invariance \cite{FAN_top_QC}. The topological aspects of quasicrystals have been recently investigated \cite{GOLDMAN, FUCHS_VIDAL, FAN_top_QC, PAI_fractal_top, BITANROY_TOP_BRANES}, but methods are often tailored to each model, and a general framework to study topology in these systems is lacking. 

Arguably the most investigated quasicrystal is the Fibonacci chain \cite{Jagannathan_FIBO_REVIEW}, a one-dimensional model based on the Su-Schrieffer-Heeger (SSH) model \cite{SSH}. The latter is a tight-binding model in which alternating weak and strong hopping parameters lead to a topological or trivial phase, depending on whether the last bond in the chain corresponds to a weak or strong hopping, respectively. The Fibonacci chain is a natural extension of the SSH model to the aperiodic domain \cite{NIU1990}, in which the weak and strong hopping parameters are distributed according to a Fibonacci word. This 1D tight-binding chain hosts many interesting properties, such as a multifractal energy spectrum and eigenstates \cite{ZHENG_FIBO_SPECTRUM, FUJIWARA_FIBO_WAVEF, MACE_MULTIFRACTAL}. In addition, it was shown to be equivalent to the Harper model \cite{KRAUS_ZILBERBERG}, from which it inherits its topological properties. In particular, a description of the system in terms of conumbers \cite{SIRE_CON} has revealed hidden symmetries in Hilbert space and allowed for a systematic prediction of the influence of random disorder based on a renormalization group (RG) scheme \cite{NIU1990}. The interpretation of the system in terms of local symmetries has also led to a more profound understanding of its physical properties \cite{RONTGEN_LRM}. 

In this paper, we go beyond the realm of dimerized models, such as the SSH and Fibonacci chain, and introduce a quantum chain based on the Tribonacci substitution. Two tight-binding chains, the hopping Tribonacci Chain (HTC) and the on-site Tribonacci Chain (OTC), are defined analogously to the Fibonacci chain. These chains are closely linked to the Rauzy fractal, a well-known compact domain with fractal boundary \cite{RAUZY}. An RG scheme for the HTC and OTC is developed along the lines proposed by Niu and Nori \cite{NIU1990}. This allows for the same interpretation of the spectrum as a multifractal set as for the Fibonacci chain \cite{ZHENG_FIBO_SPECTRUM}. The RG scheme is also used to render the HTC and OTC equivalent at the RG fixed point. We show how the Rauzy fractal orders the lattice points according to their local environment, in analogy with the conumbering scheme. Furthermore, the RG procedure provides a natural way to enumerate all structures in the Local Resonator Mode (LRM) framework \cite{RONTGEN_LRM}. Finally, we compute the multifractal dimensions of the energy spectrum and eigenstates of the HTC, and compare them with the Fibonacci chain. From these results, it can be concluded that the Tribonacci chains are critical in terms of Anderson localization.

The paper is structured as follows. In section~\ref{sec:model} we introduce the HTC, the OTC, and all elements that are needed to define the model, such as the Tribonacci word and the Rauzy fractal. Section~\ref{sec:renormalization} is devoted to the RG scheme for the HTC and OTC, and how the two models can be considered equivalent in the infinite RG limit. In Section~\ref{sec:states_rauzy}, the Rauzy fractal is proposed as the analog of conumbering for the HTC and OTC. Multifractal properties of the spectrum and wavefunction of the HTC are computed in Section~\ref{sec:multifractality}, and compared to the Fibonacci chain. Finally, the conclusion and outlook are presented in Section~\ref{sec:conclusion}.



\section{The Model}\label{sec:model}
In this section, we introduce the elements needed to define the Tribonacci chain. The main element is the Tribonacci word, which determines the quasiperiodic modulation in the tight-binding chains.

\subsection{The Tribonacci Word}

\subsubsection{The Tribonacci Sequence}
Analogous to the Fibonacci sequence, one can define the Tribonacci sequence recursively as
\begin{equation}\label{eq:tribonacci_sequence}
    T_{N+1} = T_N + T_{N-1} + T_{N-2},
\end{equation}
with initial values $T_{-2} = 0, T_{-1} = T_0 = 1$. The Tribonacci constant $\beta$, the analog of the golden ratio, is obtained as the limit
\begin{equation}
    \beta = \lim_{N \to \infty} \frac{T_{N+1}}{T_N} \approx 1.8392\dots,
\end{equation}
which is also the unique real root of the polynomial
\begin{equation}\label{eq:tribonacci_polynomial}
    P(x) = x^3 - x^2 - x - 1.
\end{equation}
The other two roots $\omega, \Bar{\omega}$ are complex and satisfy $|\omega|<1$.

\subsubsection{The Tribonacci Substitution}

The Tribonacci substitution is the substitution $\rho$ on the alphabet $\mathcal{A} = \{0, 1, 2 \}$ that reads
\begin{equation}
    \rho : 
    \begin{cases}
    0 \mapsto 01, \\
    1 \mapsto 02, \\
    2 \mapsto 0.
    \end{cases}
\end{equation}
The Tribonacci word is obtained by repeatedly applying $\rho$ to the seed $W_0 = 0$. The resulting word after $N$ applications $W_N:= \rho^N(W_0)$ is called the $N$th Tribonacci approximant. The Tribonacci word is the limit $W := \lim_{N \to \infty} W_N$. The first few approximants read
\begin{align*}
    W_0 &= 0, \\
    W_1 &= 01, \\
    W_2 &= 0102, \\
    W_3 &= 0102010, \\
    W_4 &= 0102010010201. \\
\end{align*}
An alternative way to generate the Tribonacci word is by concatenating the previous three approximants
\begin{equation}
    W_{N+1} = W_{N} W_{N-1} W_{N-2},
\end{equation}
which is reminiscent of Eq.~(\ref{eq:tribonacci_sequence}). Therefore, the Tribonacci constant is equivalently obtained by the limit
\begin{equation*}
    \beta = \lim_{N \to \infty} \frac{|W_{N+1}|}{|W_N|},
\end{equation*}
where $|\cdot|$ denotes the length of the word.

Another important tool when dealing with any substitution $\rho$ is the incidence matrix $\vb{M} = [m_{ij}]$, where $m_{ij}= |\rho(j)|_i$ and $|w|_k$ denotes the number of occurrences of the letter $k$ in the word $w$. The incidence matrix is used in the relation
\begin{equation*}
    \vb{N}^{(N+1)} = \vb{M} \cdot \vb{N}^{(N)},
\end{equation*}
where $\vb{N}^{(N)} := (|W_N|_0, |W_N|_1, |W_N|_2)^T$ is the vector that counts how often each letter occurs in the approximant $W_N$. If $\vb{M}$ has precisely one eigenvalue $\lambda$ with $|\lambda| > 1$ and all other eigenvalues have modulus strictly less than $1$, the substitution is called Pisot. The incidence matrix for the Tribonacci substitution and its characteristic polynomial read
\begin{equation}\label{eq:tribonacci_adjacency_matrix}
    \vb{M} = 
\begin{pmatrix}
1 & 1 & 1 \\
1 & 0 & 0 \\
0 & 1 & 0
\end{pmatrix}, \qquad \det( \lambda \vb{I} - \vb{M}) = \lambda^3 - \lambda^2 - \lambda - 1, 
\end{equation}
which is identical to the Tribonacci polynomial Eq.~(\ref{eq:tribonacci_polynomial}). Hence, it is immediate that the Tribonacci substitution is Pisot. The eigenvalues are $\lambda = \beta > 1$ and $\lambda = \omega, \Bar{\omega}$ where $|\omega|<1$.

One can also define the bi-infinite Tribonacci word $W|W$ in a consistent way (see Ch. 4 of Ref.~\cite{BAAKE_GRIMM_APER_ORDER_1}). Take the seed $\rho^{-1}(0)|0 = 2|0$ and apply $\sigma = \rho^3$ infinitely often to the seed. This results in the approximants $W_{3N-1} | W_{3N}$ and the limit
\begin{equation}\label{eq:bi-infinite_tribonacci_word}
    W|W := \lim_{N \to \infty} W_{3N-1} | W_{3N} = \cdots w_{-2} w_{-1} | w_0 w_1 \cdots.
\end{equation}


\subsubsection{The Rauzy Fractal}\label{sec:rauzy_fractal}
\begin{figure}[tb]
\includegraphics[width = 0.8\linewidth]{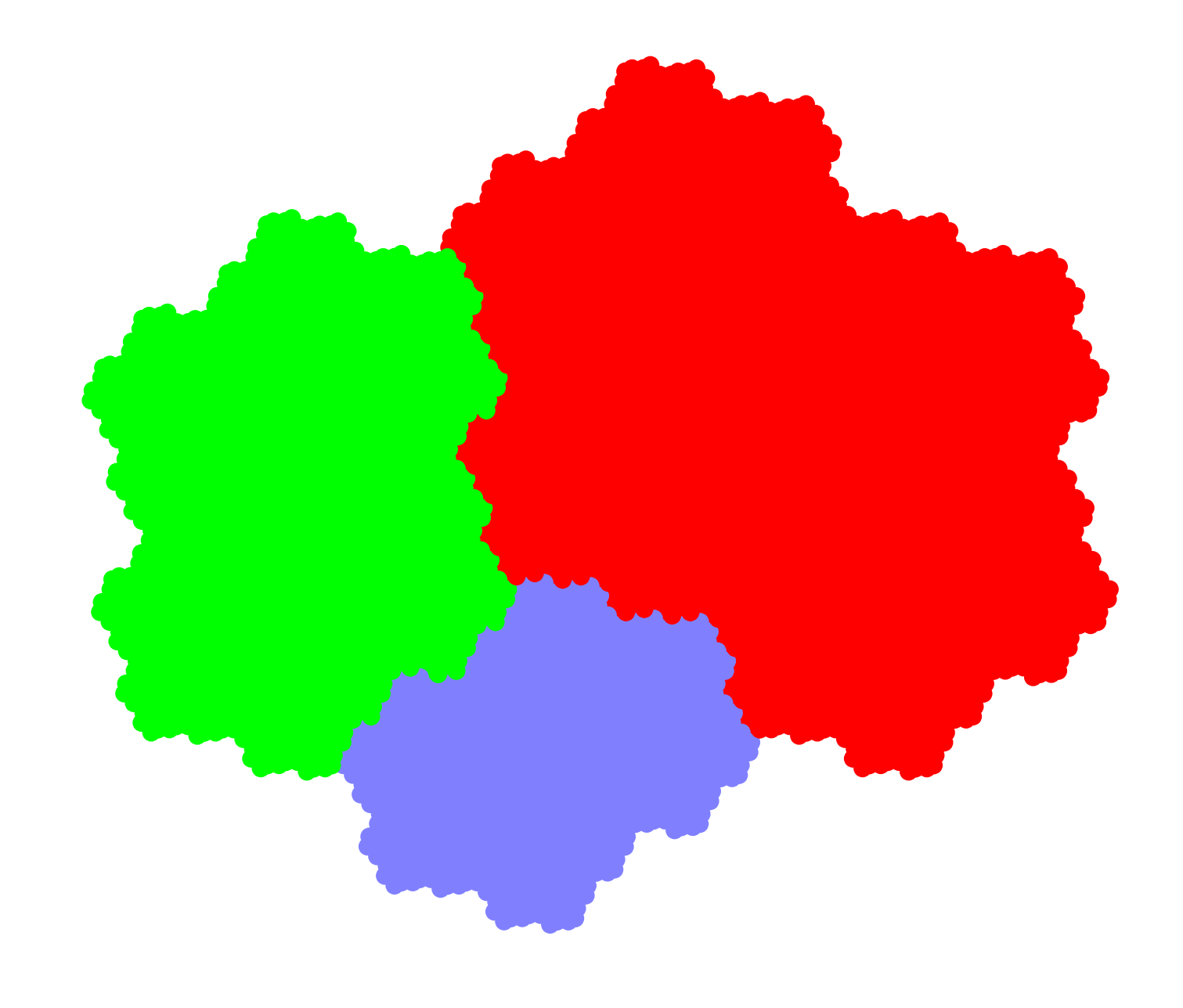}
\caption{(Color online) A Rauzy fractal with $T_{14}=5768$ points. Each region corresponds to a symbol: red (0), green (1) and blue (2). }
\label{fig:rauzy_fractal_clean}
\end{figure}

In 1982, G\'erard Rauzy used the Tribonacci substitution to define a 2D compact domain with fractal boundary, called the Rauzy fractal~\cite{RAUZY} (see Fig.~\ref{fig:rauzy_fractal_clean}). The Rauzy fractal is obtained as a subset of $\C$ via the valuation map. Let $[W]_m$ denote the first $m$ letters of the Tribonacci word and take the left eigenvector $\vb{v}=(v_0,v_1,v_2)$ of $\vb{M}$ in Eq.~(\ref{eq:tribonacci_adjacency_matrix}), corresponding to the eigenvalue $\omega$. Then, the $m$th point in the Rauzy fractal is given by
\begin{equation}\label{eq:valuation_map}
    z_m = E([W]_m) = \sum_{i \in \{0,1,2\}} |[W]_m|_i v_i \in \C,
\end{equation}
where $E$ is the valuation map and $m \geq 0$. Enumerating the letters of $W = w_0 w_1 w_2 \cdots$, each point can be assigned a color defined by the $w_m \in \{0, 1, 2 \}$, the $(m+1)$th letter~\cite{SIRVENT}. The Rauzy fractal is the compact set $\mathfrak{R} = \overline{\{z_m \mid m \geq 0 \}}$.

Another way to obtain Fig.~\ref{fig:rauzy_fractal_clean} is by starting at the origin in $\R^3$, and for each letter in $W$, taking a unit step in the $x, y$ or $z$-direction if the letter is $0$, $1$ or $2$, respectively \cite{ARNOUX_ITO}. This will create a staircase in $\R^3$ in the direction of $\vb{v}_\beta = (\beta^2,\beta,1)^T$, which spans the expanding eigenspace $L$ of Eq.~(\ref{eq:tribonacci_adjacency_matrix}). Denote $\pi_\text{int}$ the projection along $\vb{v}_\beta$ onto the 2D contracting eigenspace. Then, the $m$th point for $m \geq 0$ is given by
\begin{equation}\label{eq:rauzy_project_x_m}
    \vb{x}_m = \pi_\text{int} \sum^{m-1}_{i=0} \vb{e}_{w_i} \in \R^2,
\end{equation}
where $\vb{e}_i$ are the canonical basis vectors of $\R^3$. The Rauzy fractal is the compact set $\mathfrak{R}' = \overline{\{ \vb{x}_m \mid m \geq 0 \}}$, which is not precisely $\mathfrak{R}$, but related by an affine transformation (see Appendix~\ref{app:rauzy} for details of this transformation).


\subsubsection{Cut-and-Project Sets}

Bearing the Rauzy fractal $\mathfrak{R}'$ in mind, one can view the Tribonacci word as a quasicrystal. Consider again the Tribonacci staircase, which are the points $\vb{y}_m = \sum^{m-1}_{i-0}\vb{e}_{w_i}$. Using the bi-infinite word $W|W$, the staircase can also be defined for $m<0$ by $\vb{y}_m = \sum^{-1}_{i=m} \vb{e}_{w_i}$. From the bi-infinite staircase, one can construct a 1D Tribonacci quasicrystal
\begin{equation}\label{eq:tribo_lattice}
    \Lambda_\text{trib} = \{ \pi \vb{y}_m \mid m \in \Z \},
\end{equation}
by projecting all staircase points along the stable eigenspace onto the line spanned by $\vb{v}_\beta$, where this projection is denoted $\pi$. 

One can see that $\Lambda_\text{trib}$ is a cut-and-project set in the following sense. Take a cubic lattice in $\R^3$ and trace out a volume by sliding the set $\mathfrak{R}'$, the acceptance set of the cut-and-project scheme, along the space $L$. Note that all lattice points lying in the traced out volume are exactly the staircase points $\vb{y}_m$, which constitute $\Lambda_\text{trib}$ upon projecting onto $L$. A key result is that any cut-and-project set has a point diffraction pattern \cite{BAAKE_GRIMM_APER_ORDER_1}, which leads to the conclusion that the aperiodic lattice $\Lambda_\text{trib}$ is a quasicrystal.

Finally, we would like to point out that there exists a quasiperiodic 2D tiling, the Rauzy tiling, which is based on the Tribonacci numbers and is cut-and-project set from a 3D space \cite{VIDAL_2001_GEN_RAUZY_TILINGS}. Several physical properties of tight-binding models on these lattices have been studied \cite{JAGANNATHAN_RAUZY2001, VIDAL_RAUZY2002}, in particular the effect of a magnetic field \cite{VIDAL2004, GOLDMAN, FUCHS_VIDAL}. The generalized Rauzy tiling extends this construction to arbitrary dimension, and this family of tilings can be viewed as a generalization of the Fibonacci chain \cite{VIDAL_2001_GEN_RAUZY_TILINGS}.

\subsubsection{Recurrence Properties}\label{sec:tribo_recurrence}

Another key property of the Tribonacci $W$ word is its self-similarity \cite{BERTHE}. Take any finite string $s=s_1 \cdots s_N$ of length $N$ that occurs somewhere in $W$. We say that $s$ occurs at position $i$ in $W$ if $s_1 \cdots s_N = w_i \cdots w_{i+N}$. Let $i_1,i_2,\dots$ denote the places where $s$ occurs in $W$. Then, the words $r_j = w_{i_j} \cdots w_{i_{j+1}}$ between occurrences of $s$ have useful properties. Firstly, for any choice $s$, the word $r_j \in \{r^{(0)}, r^{(1)}, r^{(2)}\}$ takes one of three values. Secondly, if we label $r^{(i)}$ such that $r^{(0)}$ occurs most often, $r^{(1)}$ second most often and $r^{(2)}$ least often, then the map $\kappa : r^i \mapsto i$ maps the string $r_1 r_2 \cdots$ back to $W$. In other words
\begin{equation}\label{eq:tribo_word_renorm}
    \kappa(r_1) \kappa(r_2) \cdots = W,
\end{equation}
where $r_i$ are the words between subsequent occurrences of $s$ in $W$. This also works if $s$ occurs in a Tribonacci approximant $W_N$. By applying periodic boundary conditions when determining $r_j$, the map $\kappa$ results in
\begin{equation}\label{eq:tribo_word_renorm_finite}
    \kappa(r_1) \kappa(r_2) \cdots = W_{N - k},
\end{equation}
where $k$ depends on the choice of $s$. Eqs.~(\ref{eq:tribo_word_renorm}) and (\ref{eq:tribo_word_renorm_finite}) are the foundation of the perturbative RG scheme in Section~\ref{sec:renormalization}.


We would like to emphasise that there are other quantum chains based on three-letter substitutions that are Pisot with the same dominant $\lambda = \beta$. One such example is the system studied by Ali et al. \cite{ALI_GUMBS}. This is fundamentally different from our work, since in their case there is not a natural RG scheme and our connection to the Rauzy fractal is entirely new.

\subsection{Tribonacci Tight-Binding Models}
The definition of the Tribonacci chain, with aperiodic hopping and on-site energy, generalizes the work by Niu and Nori~\cite{NIU1990} on the Fibonacci chain to the HTC and OTC.

\subsubsection{Hopping Model}
The infinite HTC is defined as a 1D tight-binding chain with no on-site potentials and hopping parameters that are modulated according to the Tribonacci word $W|W$. The Hamiltonian reads
\begin{equation}\label{eq:infinite_tribonacci_chain_hamiltonian}
    H = \sum_{n \in \Z} t_{w_n} \ket{n+1} \bra{n} + H.c.,
\end{equation}
where $w_n \in \{0,1,2\}$ are the letters of $W|W$ in Eq.~(\ref{eq:bi-infinite_tribonacci_word}) and the model is parameterized by one parameter $\rho \in [0,1]$ as $t_0 / t_1 = t_1 / t_2 = \rho$. Note that Eq.~(\ref{eq:infinite_tribonacci_chain_hamiltonian}) possesses chiral symmetry $\Gamma H \Gamma = -H$, where $\Gamma^2 = 1$ and
\begin{equation*}
    \Gamma = \sum_{n \in Z} \ket{2n} \bra{2n} - \sum_{n \in Z} \ket{2n+1} \bra{2n+1}.
\end{equation*}
A direct consequence of chiral symmetry is a symmetric spectrum around $E=0$. The model is studied in the regime where $\rho \ll 1$, i.e. $0<t_0 \ll t_1 \ll t_2$, such that there is a hierarchy of bond strengths, analogous to Ref.~\cite{NIU1990}.

\subsubsection{On-Site Model}
The OTC is defined by the Hamiltonian 
\begin{equation}\label{eq:infinite_onsite_tribonacci_chain_hamiltonian}
    H = \sum_{n \in \Z}  \epsilon_{w_n} \ket{n}\bra{n} -t \sum_{n\in \Z} \ket{n+1} \bra{n} + H.c. , 
\end{equation}
where now the hopping parameters $t$ are constant, and the on-site potential $\epsilon_i$ is modulated according to the Tribonacci word $W|W$. This model is generally parameterized by two parameters, $c_1 = (\epsilon_1-\epsilon_0)/t$ and $c_2 = (\epsilon_2 - \epsilon_0)/t$. Analogous to Ref.~\cite{NIU1990}, we demand $|c_1|,|c_2|,|c_2-c_1| \gg 1$, which physically means that the on-site potentials dominate and are weakly coupled. One particular choice is $c_1 = c_2/2 = c \gg 1$, which will be used when comparing to the HTC.

\section{Perturbative Renormalization of the Tribonacci Chain}\label{sec:renormalization}
We now present the perturbative RG scheme for the HTC and OTC. The scheme is possible due to the self-similar recurrence properties of the Tribonacci word (see Section~\ref{sec:tribo_recurrence}), and is analogous to the RG of the Fibonacci chain proposed by Niu and Nori \cite{NIU1990} (see the review by Jagannathan \cite{Jagannathan_FIBO_REVIEW} for more details on the Fibonacci chain). 


\subsection{The Renormalization Scheme}

\subsubsection{Hopping Model}\label{sec:renorm_hopping}
For the RG scheme, it is convenient to consider the $N$th HTC approximant
\begin{equation}\label{eq:tribonacci_chain_finite_periodic}
    H_N = \sum_{n=0}^{T_N-1} t_{w_n} \ket{n+1 \bmod T_N} \bra{n} + H.c.,
\end{equation}
where periodic boundary conditions are enforced. Furthermore, the Hamiltonian is split up in two parts
\begin{equation}\label{eq:H_N_split}
    H_N = H_{0,N} + H_{1,N},
\end{equation}
where $H_{1,N}$ contains only the terms with a $t_0$ hopping, such that $H_{0,N}$ can be regarded as the unperturbed Hamiltonian. Note that $H_{0,N}$ has only five highly degenerate energy levels $E = 0, \pm t_1, \pm t_2$. The $E=0$ states are the atoms, which are isolated sites, corresponding to $00$ in $W$. Type-1 molecules are the $E=\pm t_1$ states, corresponding to $010$ in $W$. These are isolated dimers consisting of two neighboring sites, coupled by a $t_1$ bond, which can either bond or anti-bond. Similarly, the $E = \pm t_2$ states correspond to $020$ in $W$, and are called type-2 molecules. 

\begin{figure}[t]
\includegraphics[width = \linewidth]{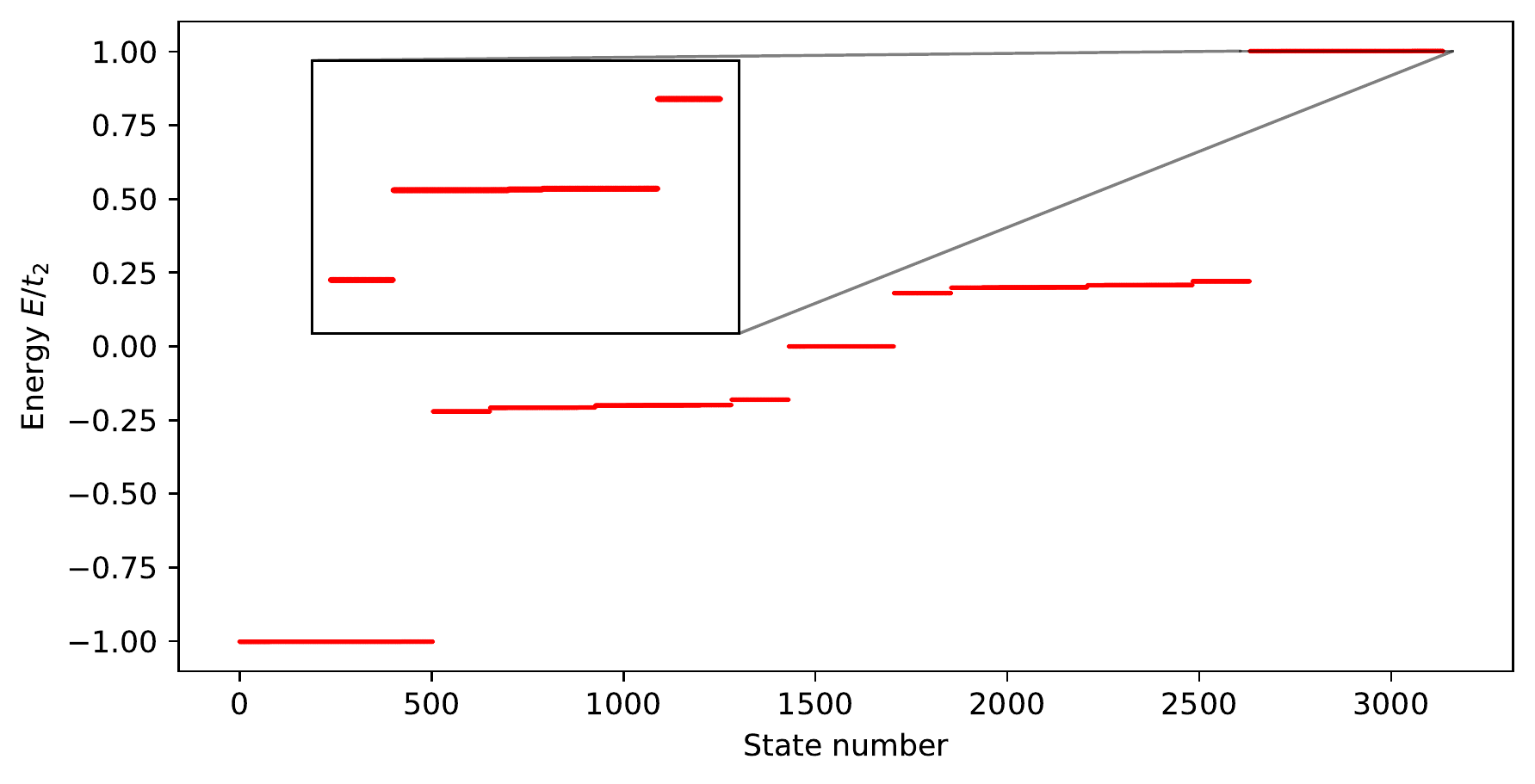}
\caption{(Color online) The energy spectrum of the HTC Eq.~(\ref{eq:tribonacci_chain_finite_periodic}) with $\rho=0.2$ and $T_{13}=3136$ sites. The five main bands are located around $E=0,\pm t_1, \pm t_2$. The inset shows a zoom in on the top band, which exhibits a similar spectrum, but with seemingly different $t_0,t_1$.} 
\label{fig:tribo_chain_spectrum}
\end{figure}

Upon setting $t_0$ nonzero, the atoms/molecules start to interact. If one considers one type of atom or molecule as a lattice site, one can compute the effective coupling between subsequent sites using Brillouin-Wigner perturbation theory. Fig.~\ref{fig:tribo_chain_spectrum} depicts the spectrum of Eq.~(\ref{eq:tribonacci_chain_finite_periodic}), where one can see five branches around $E = 0, \pm t_1, \pm t_2$ that would become fully degenerate upon setting $t_0=0$.

Now, we explain the simplest case, the type-1 molecule, in detail. The procedure for the other bonds is exactly the same, but with longer computations. Consider the Tribonacci approximant 
\begin{multline}\label{eq:tribo_approx_W6}
        W_6 = \\
\textcolor{red}{0} 1 \overset{r_1}{\underline{\textcolor{red}{020}}} 1 \overset{r_2}{\textcolor{red}{00}}1\overset{r_3}{\textcolor{red}{020}}\overset{r_4}{1\textcolor{red}{0}1}\overset{r_5}{\textcolor{red}{020}}1\overset{r_6}{\textcolor{red}{00}}1\overset{r_7}{\textcolor{red}{020}}1\overset{r_8}{\textcolor{red}{020}}1\overset{r_9}{\textcolor{red}{00}}1\overset{r_{10}}{\textcolor{red}{020}}\overset{r_{11}}{1\textcolor{red}{0}1}\overset{r_{12}}{\textcolor{red}{020}}1 \overset{r_{13}}{\textcolor{red}{0}. \,} 
\end{multline}
The first step is to tabulate all words $r_i$ occurring between $1$'s in $W_6$, starting after the first occurrence of $1$, and considering periodic boundary conditions. The possibilities are $020, 00$ and $0$, which occur $7, 4$ and $2$ times, respectively. Therefore
\begin{equation}\label{eq:tribo_approx_rs}
\begin{gathered}
    \{ r \} =  \{ r_1 = r^{(0)}, r_2 = r^{(1)}, r_3 = r^{(0)}, \dots, r_{13} = r^{(1)} \}, \\ 
    r^{(0)} = 020, r^{(1)} = 00, r^{(2)} = 0. 
\end{gathered}
\end{equation}
Finally, upon applying the map $\kappa : r^{(i)} \mapsto i$, the Tribonacci approximant $W_4$ is obtained as
\begin{equation}\label{eq:tribo_approx_W4}
    W_4 =  \kappa(r_1) \kappa(r_2) \cdots \kappa(r_{13}) =  0102010010201,
\end{equation}
which has $k=2$ in Eq.~(\ref{eq:tribo_word_renorm_finite}). The procedure in Eqs.~(\ref{eq:tribo_approx_W6}), (\ref{eq:tribo_approx_rs}), and (\ref{eq:tribo_approx_W4}), which is the $s=1$ case, can be carried out for any $s$. This is done for $s=0,1,2,00$ in Table~\ref{tab:tribo_renorm_words}. 

\begin{table}[tb]
\caption{\label{tab:tribo_renorm_words} For $W_N$ and particular strings $s=0,1,2,00$, the occurrences between $s$ can be one of $r^{(i)}$, and map to $W_{N-k}$ under the map $\kappa: r^{(i)} \mapsto i$, $i=0,1,2$.}
\centering
\begin{tabular}{||c | c c c c||} 
 \hline
  $s$ & $r^{(0)}$ & $r^{(1)}$ & $r^{(2)}$ & maps $W_N$ to \\ [0.5ex] 
 \hline\hline
 0 & 1 & 2 & $\emptyset$ & $W_{N-1}$   \\ 
 \hline
 1 & 020 & 00 & 0 & $W_{N-2}$   \\
 \hline
 2 & 010010 & 01010 & 010 & $W_{N-3}$  \\
 \hline
 00 & 10201010201 & 102010201 & 10201 & $W_{N-4}$   \\ 
 \hline
\end{tabular}
\end{table}

\begin{figure}[tb]
\includegraphics[width = \linewidth]{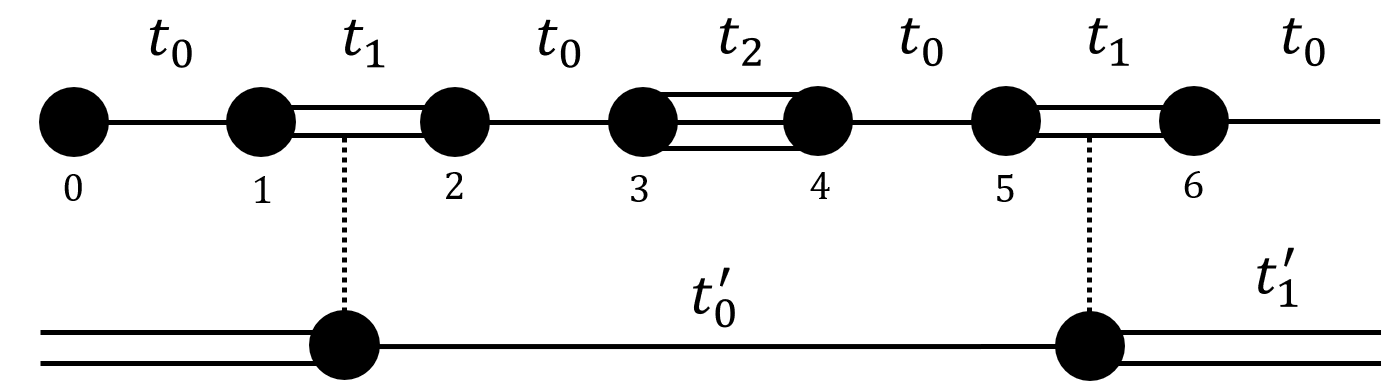}
\caption{The 3rd approximant HTC, with Hamiltonian $H_3$ (see Eq.~(\ref{eq:tribonacci_chain_finite_periodic})) and periodic boundary conditions. The single line denotes a $t_0$ bond, the double line a $t_1$ bond and a triple line a $t_2$ bond. The chain is renormalized by considering the type-1 molecules as the new lattice sites, and the chain between these molecules as the new bonds, which are $t_0', t_1'$. The figure is inspired by Ref.~\cite{NIU1990}.} 
\label{fig:tribo_renorm_chains}
\end{figure}

The procedure outlined in Eqs.~(\ref{eq:tribo_approx_W6}), (\ref{eq:tribo_approx_rs}) and (\ref{eq:tribo_approx_W4}) is applied to the HTC Hamiltonian in Eq.~(\ref{eq:tribonacci_chain_finite_periodic}) as follows. Consider the approximant in Fig.~\ref{fig:tribo_renorm_chains}. Each dimer of two sites coupled by $t_1$ is considered a lattice site in the renormalized chain, on which a (anti-)bonding state $\ket{\pm}_i$ can sit. Using perturbation theory, the effective coupling between neighboring sites
\begin{equation*}
    t'_i = \bra{\pm}_i H_{1,N} \ket{\pm}_{i+1},
\end{equation*}
is computed. The perturbation theory framework is explained in Appendix~\ref{App:BWPT}, for the chain shown in Fig.~\ref{fig:tribo_renorm_chains}. The computations and results for the hopping and on-site chain are presented in Appendix~\ref{App:BWPT_hopping} and \ref{App:BWPT_onsite}, respectively.  

The main result of the RG scheme can now be stated. Denote $H^{(p,q)}_N$ given by Eq.~(\ref{eq:tribonacci_chain_finite_periodic}) with $t_0 /t_1 = \rho^p,t_1/t_2 = \rho^q$, where $p,q \in [0,\infty)$. Setting $t_0=0$, the HTC Hamiltonian $H_{0,N}$ has $T_{N-3}, T_{N-2}, T_{N-4}, T_{N-2}$ and $T_{N-3}$ states with $E = - t_2, E= - t_1, E=0, E=t_1$ and $E=t_2$, respectively. To each of these five energies, we associate an atomic ($s=00$), bonding or anti-bonding chain ($s=1,2$). The result of the perturbative calculation (see Appendix~\ref{App:BWPT_hopping}) is
\begin{widetext}
\begin{equation}\label{eq:tribo_renorm_result}
    H_N^{(p,q)} \approx (z_2 H^{(p+q,p+2q)}_{N-3} - t_2) \oplus (z_1 H_{N-2}^{(q,p)} - t_1) \oplus (z_0 H^{(p,2p+q)}_{N-4}) \oplus (z_1 H_{N-2}^{(q,p)} + t_1) \oplus (z_2 H^{(p+q,p+2q)}_{N-3} + t_2),
\end{equation}
\end{widetext}
where the parameters read $z_0 = \rho^{4p + 2q}, z_1 = \rho^{p+q}/2$ and $z_2 = \rho^{2p+3q}/2$. The computation of the parameters $z_i$ and the $p,q$ exponents in each of the five blocks is identical to Ref.~\cite{NIU1990}, and is repeated in detail in Appendix~\ref{App:BWPT_hopping}. The HTC in Eq.~(\ref{eq:tribonacci_chain_finite_periodic}) realizes the case $p=q=1$. From the result Eq.~(\ref{eq:tribo_renorm_result}), it is clear that the spectrum consists not simply of scaled and shifted versions of itself, but rather related spectra of chains with various $p,q$ values. Since one can identify each of the five quasibands in Fig.~\ref{fig:tribo_chain_spectrum} to a block in Eq.~(\ref{eq:tribo_renorm_result}), the spectrum can be interpreted as a multifractal set as Zheng \cite{ZHENG_FIBO_SPECTRUM} did for the Fibonacci chain.

The words $r^{(i)}$ in Table~\ref{tab:tribo_renorm_words} are longer than those in the RG scheme for the Fibonacci chain, which requires higher orders of perturbation theory to yield a nonzero result. This has the advantage that the error made in the approximate RG Eq.~(\ref{eq:tribo_renorm_result}) is smaller than the RG scheme for the Fibonacci chain.


\subsubsection{On-Site Model}
\begin{figure}[b]
\includegraphics[width = \linewidth]{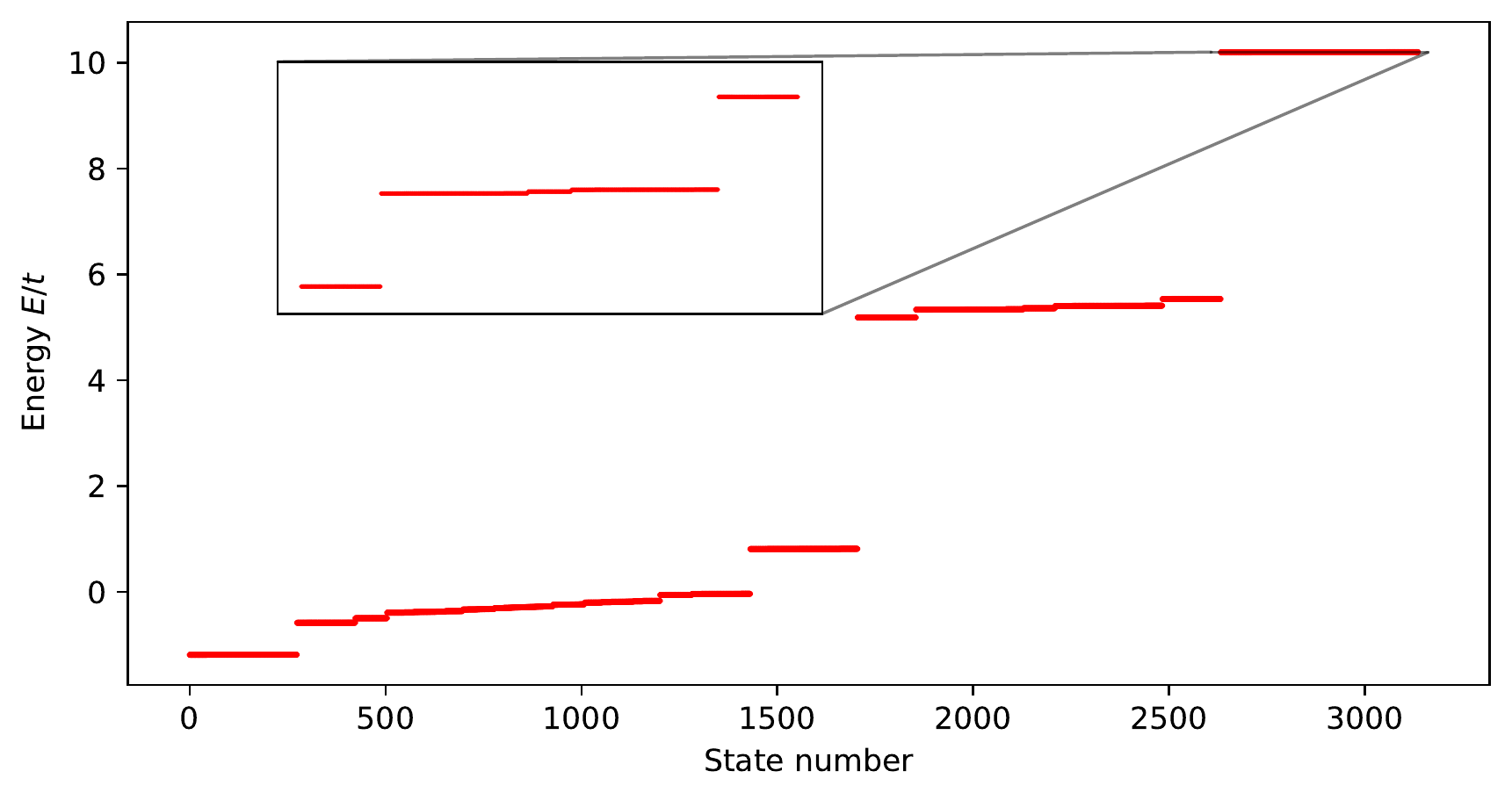}
\caption{(Color online) The energy spectrum of the OTC Eq.~(\ref{eq:tribonacci_chain_finite_periodic_onsite}) with $c=5$ and $T_{13}=3136$ sites. The three main bands sit around $E/t = 0, c, 2c$. The inset shows that the main bands further split up like HTC Hamiltonians with certain $p,q$ values.} 
\label{fig:tribo_chain_spectrum_onsite}
\end{figure}

The $N$th approximant of the OTC is defined as 
\begin{equation}\label{eq:tribonacci_chain_finite_periodic_onsite}
    H_N^{o} = \sum^{T_N-1}_{n=0} \epsilon_{w_n} \ket{n} \bra{n} - t\big(\ket{n+1 \bmod T_N} \bra{n} + H.c. \big),
\end{equation}
where periodic boundary conditions are enforced. When writing
\begin{equation}
    H_N^o = H_{0,N}^o + H_{1,N}^o,
\end{equation}
the part $H_{1,N}^o$ consists of all $t$ bonds and $H_{0,N}^o$ only the on-site energies. At $t=0$, the chain consists of $T_{N-1}, T_{N-2}$ and $T_{N-3}$ isolated sites with energy $E = 0, \epsilon_1, \epsilon_2$, respectively. When $t$ is nonzero, the degeneracy is lifted and the spectrum consists of three bands, as depicted in Fig.~\ref{fig:tribo_chain_spectrum_onsite}.

The analysis in Section~\ref{sec:renorm_hopping} can be immediately carried over to the three atomic chains of the on-site model, to approximate each of the three bands as a general HTC $H^{(p,q)}_{N-k}$. For a general OTC parameterized by $c_1$ and $c_2$, the result of the perturbation theory (see Appendix~\ref{App:BWPT_hopping}) reads
\begin{widetext}
\begin{equation}\label{eq:tribo_renorm_result_onsite}
    H_N^o \approx (z_0 H_{N-1}^{(p_0,q_0)}+\epsilon_0) \oplus (z_1 H_{N-2}^{(p_1,q_1)}+\epsilon_1) \oplus (z_2 H_{N-3}^{(p_2,q_2)}+\epsilon_2),
\end{equation}
\end{widetext}
where $z_0 = t, z_1 = t/c_1, z_2 = t/[c_2^2(c_2-c_1)]$ and $p_i = \log a_i / \log \rho, q_i = \log b_i / \log \rho$ for $i=0,1,2$ where $a_i,b_i$ are given in Table~\ref{tab:tribo_onsite_to_hopping} in the appendix.

As a final remark, by the recurrence property Eq.~(\ref{eq:tribo_word_renorm}) of the infinite word $W$, the approximations Eqs.~(\ref{eq:tribo_renorm_result}) and (\ref{eq:tribo_renorm_result_onsite}) are also valid in the infinite limit, where the subscripts $N, N-k$ are dropped.

\subsection{Hopping and On-Site Equivalence}\label{sec:equivalence}

The RG scheme of the HTC Eq.~(\ref{eq:tribo_renorm_result}) can be repeatedly applied to the five HTC Hamiltonians in the direct sum. For the OTC, the same is true after one application of the RG Eq.~(\ref{eq:tribo_renorm_result_onsite}). Considering the infinite HTC and OTC, the Hamiltonian after $m$ applications of the RG is described by $5^m$ and $3 \cdot 5^{m-1} $ pairs of $p,q$ values, respectively. We will show that the HTC and OTC are equivalent, in the sense that for both models, the fraction of $p,q$ values that escape to infinity tends to one as $m \to \infty$.

For the HTC, define $I_m = \{ p_i,q_i \mid i=1,\dots,5^m \}$, the set of $p,q$ values in the direct sum after $m$ RG applications. Define the probability measure on the measurable space $(I_m, 2^{I_m})$ as
\begin{equation}
    \mu_m (A) := |A|/|I_m|,
\end{equation}
where $2^{I_m}$ denotes the powerset, $A \subset I_m$ and $|\cdot|$ denotes the cardinality of the set. To study the divergence of $p,q$ values, define the set of values smaller than $m$ as
\begin{equation}
    J_m := \{x \in I_m \mid x \leq m \}.
\end{equation}
For the OTC, all objects $I^o_m,J^o_m,\mu^o_m$ are similarly defined.

The mathematical statement of the equivalence, as proven in Appendix \ref{app:equivalence_TC_OTC}, reads
\begin{equation}\label{eq:tribo_equiv_goal}
    \lim_{m \to \infty} \mu_m(J_m) = \lim_{m \to \infty} \mu_m^o(J^o_m) = 0.
\end{equation}
This proves that for both the HTC and OTC, the set of $p$ and $q$ values that remain finite can be at most a set of measure zero. This means that both the HTC and OTC are described by an infinite direct sum of $H^{(p,q)}$ Hamiltonians with $p=q=\infty$, which are Tribonacci chains where only the $t_2$ bonds are nonzero.

The similarity discussed in this work is a different notion of similarity than Niu and Nori \cite{NIU1990} proved for the Fibonacci chain. In their case, all values would read $p=1$ and $q=1$ in Eqs.~(\ref{eq:tribo_renorm_result}) and (\ref{eq:tribo_renorm_result_onsite}). Since the Fibonacci chain perturbatively renormalizes to exact scaled copies of itself, it can be viewed as a critical model. The Tribonacci chains renormalize perturbatively to different kinds of HTCs, viz. HTCs with $p\neq 1$ and/or $q\neq 1$.  The limit of the RG procedure for the HTC yields infinitely many copies of the HTC with $p=q=\infty$, which is quite different from the original model where $p=q=1$. In this way, the HTC and OTC can be viewed as less critical than the Fibonacci chain. Regardless of this fact, in Section~\ref{sec:multifractality} it will be shown that the eigenstates of the HTC are critical.

\section{Eigenstates on the Rauzy Fractal}\label{sec:states_rauzy}







Considering the Hamiltonian $H_N$ in Eq.~(\ref{eq:tribonacci_chain_finite_periodic}) (or Eq.~(\ref{eq:tribonacci_chain_finite_periodic_onsite})), the Schr\"odinger equation $H_N \ket{\psi}_i = E_i \ket{\psi}_i$ will have $T_N$ solutions labeled by $i=0,\dots,T_N-1$. Each eigenstate has the form $\ket{\psi}_i = \sum^{T_N-1}_{n=0} \psi_i(n) \ket{n}$, where $\psi_i(n) \in \C$. The eigenstate $\ket{\psi}_i$ can be plotted on the Rauzy fractal by identifying each point $\vb{x}_n$ in Eq.~(\ref{eq:rauzy_project_x_m}) with the probability $|\psi_i(n)|^2$, which determines the size of a black triangle at that point.

\subsection{Hopping Model}

When associating HTC lattice points $\ket{n}$ with Rauzy fractal points $\vb{x}_n$, one has to apply a different coloring of the Rauzy fractal. Each site has no on-site energy, and can have the local environments 
\begin{itemize}
    \item Red: 01 ($t_0$ on the left and $t_1$ on the right) or 10,
    \item Green: 02 or 20,
    \item Blue: 00.
\end{itemize}

\begin{figure*}
    \includegraphics[width = \textwidth]{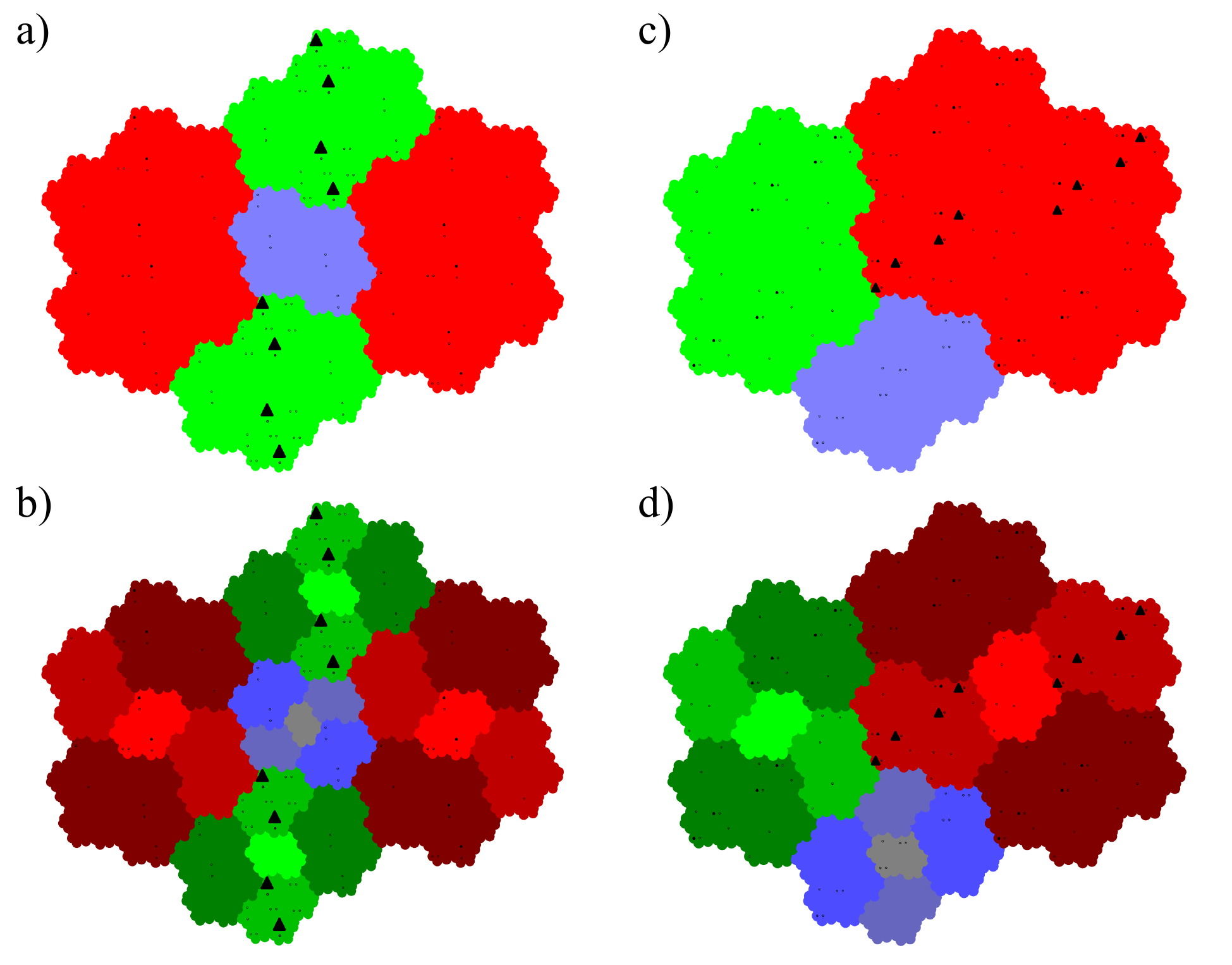}
    \caption{(Color online) The eigenstate $\ket{\psi}_0$ on the Rauzy fractal of $T_{13}=3136$ points. The regions are colored according to the local environment of a lattice site $n$ in the HTC (or OTC), and the length of the black triangles are proportional to $|\psi_0(n)|^2$. (a) $\ket{\psi}_0$ of the HTC $H_{13}$ with coupling $\rho=0.2$ and coloring according to nearest-neighbor bonds. (b) $\ket{\psi}_0$ of the HTC $H_{13}$ with coupling $\rho=0.2$ and coloring according to the five possible environments of the local structures in a). (c) $\ket{\psi}_0$ of the OTC $H_{13}^o$ with coupling $c=5$ and and coloring according the on-site potential of a lattice site. (d) $\ket{\psi}_0$ of the OTC $H_{13}^o$ with coupling $c=5$ and coloring according to the five possible environments of the lattice sites in c).} 
\label{fig:tribo_rauzy_abcd_subdiv}
\end{figure*}

The eigenstate $\ket{\psi}_0$ of the HTC $H_{13}$ is plotted on the Rauzy fractal in Fig.~\ref{fig:tribo_rauzy_abcd_subdiv}(a). Since the energy $E_0$ comes from the bottom branch of the spectrum in Fig.~\ref{fig:tribo_chain_spectrum}, it should be a state that antibonds on sites connected with $t_2$ bonds. This is precisely reflected by the plot on the Rauzy fractal in Fig.~\ref{fig:tribo_rauzy_abcd_subdiv}(a), since the eigenstate is mainly localized in the green region, corresponding to a site neighboring a $t_2$ bond. Generally, a state from branch $1$ (or $2,3,4,5$), in this case from the lowest set of eigenvalues at $E=-t_2$, in Fig.~\ref{fig:tribo_chain_spectrum} is primarily localized in the green (or red, blue, red, green) region(s) in Fig.~\ref{fig:tribo_rauzy_abcd_subdiv}(a) (see Appendix~\ref{app:eigenstates} for more examples). Finally, for any $H_N$, each red, green and blue region contains exactly $T_{N-2}, T_{N-3}$ and $T_{N-4}$ points, matching the amount of points in each branch of the spectrum.
 


For each local structure, there are again exactly five distinct environments around that structure. For example, the environment of $01$ or $10$ is always $x010y$ where $xy = 02,20,21,12$ or $22$. It turns out that these correspond exactly to the local structures $01,10,02,20,00$ of the type-1 molecular chain. The subdivisions of the Rauzy fractal are carried out in Fig.~\ref{fig:tribo_rauzy_abcd_subdiv}(b). 

We have shown that if one is interested in all possible environments of a lattice site, it is enough to consider only the nearest-neighbor environments and the RG scheme. Using the RG scheme, next variations on the nearest-neighbor environments of a lattice site are given by the nearest-neighbor environments of the renormalized chain to which that lattice site belongs.



\subsection{On-Site Model}

When plotting the eigenstates of the OTC onto the Rauzy fractal, the original coloring can be used, since each lattice site $\ket{n}$ corresponds to some $\epsilon_{w_n}$. The state with index $i=0$ is plotted on the Rauzy fractal in Fig.~\ref{fig:tribo_rauzy_abcd_subdiv}(c). Since $E_0$ comes from the bottom branch of the spectrum in Fig.~\ref{fig:tribo_chain_spectrum_onsite}, the eigenstate is localized on the red part of the Rauzy fractal which corresponds to $0$ in $W$. It is again a general feature that states from some branch in the spectrum localize on the corresponding part of the Rauzy fractal (see Appendix~\ref{app:eigenstates} for more examples). 

Since the on-site model renormalizes to three hopping models in Eq.~(\ref{eq:tribo_renorm_result_onsite}), additional subdivision of the Rauzy fractal based on next local environments yield a similar subdivision as for the hopping model. This is displayed in Fig.~\ref{fig:tribo_rauzy_abcd_subdiv}(d).

We would like to point out the similarity between the eigenstates $\ket{\psi}_0$ in Fig.~\ref{fig:tribo_rauzy_abcd_subdiv}(a) and in the red region in Fig.~\ref{fig:tribo_rauzy_abcd_subdiv}(d). This can be understood by the fact that the eigenstate $\ket{\psi}_0$ of the OTC $H^o_{13}$ is approximately the eigenstate of the first block of Eq.~(\ref{eq:tribo_renorm_result_onsite}), which is a HTC.

Another observation is the self-similar structure of the eigenstates on the Rauzy fractals in Fig.~\ref{fig:tribo_rauzy_abcd_subdiv}. This is a signature of critical eigenstates \cite{EVERS_ANDERSON}, which are also characteristic of the Fibonacci chain \cite{Jagannathan_FIBO_REVIEW}. For the Tribonacci chains, fractality is discussed in Section~\ref{sec:multifractality}.

\subsection{Equivalence Local Environment and Local Resonator Modes}\label{sec:LRM_RG_equivalence}

It is an interesting fact that all local environments are known from only the nearest-neighbor structures and the RG Eq.~(\ref{eq:tribo_renorm_result}). This fact can be applied to elegantly categorize all LRMs of the HTC and OTC. This LRM framework was developed by R\"ontgen et al. \cite{RONTGEN_LRM}, and applied to the Fibonacci chain. 

\begin{figure}[htb]
\includegraphics[width = \linewidth]{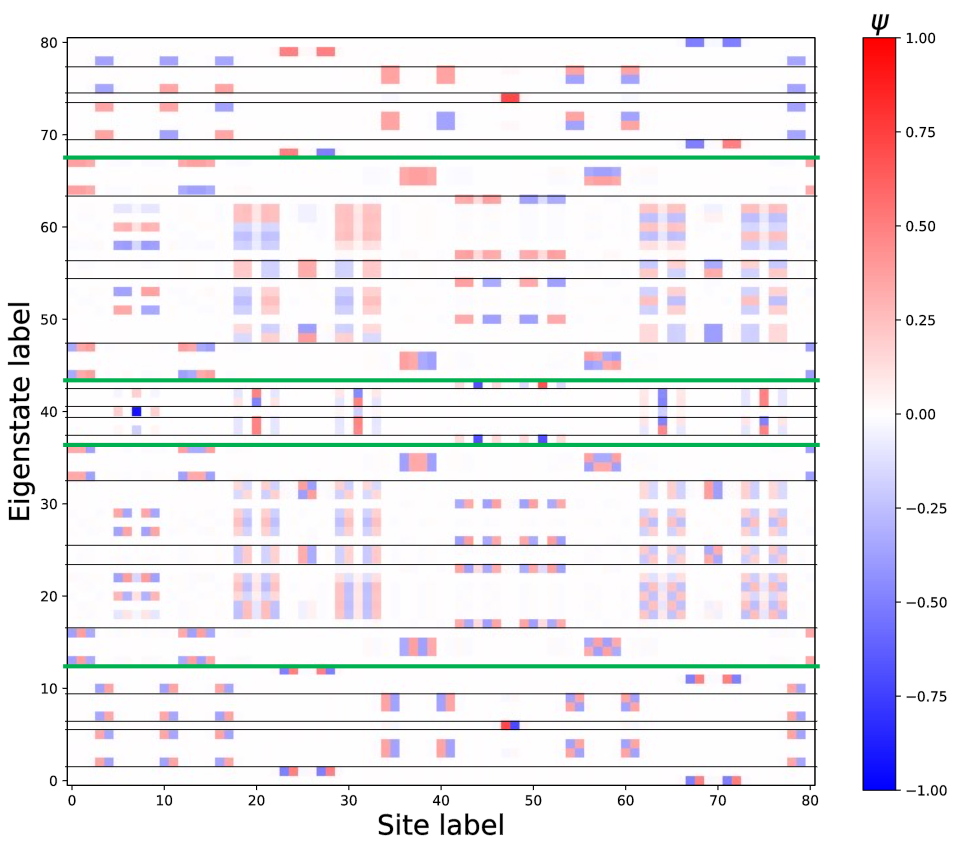}
\caption{(Color online) The HTC eigenstates $\ket{\psi}_i$ of $H_7$, ordered such that $E_i<E_{i+1}$. The sign and magnitude on each site is represented by a color. The green lines denote the splitting after one RG step, the black lines denote two RG steps. Note that the states between two subsequent lines localize on similar local environments, which is more accurate for the black lines than for the green lines.} 
\label{fig:LRM_hopping}
\end{figure}

\begin{figure}[htb]
\includegraphics[width = \linewidth]{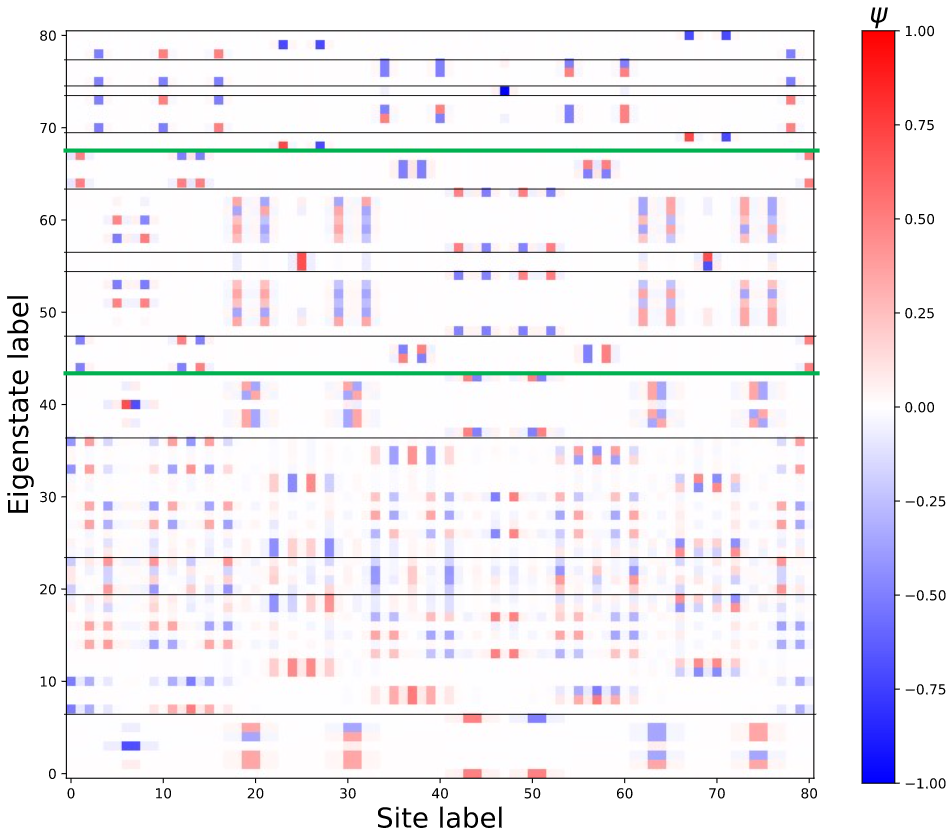}
\caption{(Color online) The OTC eigenstates $\ket{\psi}_i$ of $H_7^o$. The colors and green/black lines have the same meaning as in Fig.~\ref{fig:LRM_hopping}.} 
\label{fig:LRM_onsite}
\end{figure}

In Figs.~\ref{fig:LRM_hopping} and \ref{fig:LRM_onsite}, the eigenstate magnitude on each lattice site with $T_7=81$ sites is plotted for every energy level. The green lines define regions that precisely correspond to the diagonal blocks in Eqs.~(\ref{eq:tribo_renorm_result}) and (\ref{eq:tribo_renorm_result_onsite}), so they correspond to one application of the RG scheme. By applying Eq.~(\ref{eq:tribo_renorm_result}) again to each of these blocks of the Hamiltonian at hand, the regions subdivide again into five smaller ones (see black horizontal lines). The connection with the LRM framework is that the subsequent subdivisions order the eigenstates according to their local structure, i.e. where they are mostly localized. This classification is an essential step in the application of the LRM framework, which is a naturally carried out by the RG equations.

The RG scheme naturally gives all environments of a lattice site, and at the same time categorizes the LRMs. This simplification of the analysis is founded on the self-similarity of the Tribonacci word.

\section{Multifractality}\label{sec:multifractality}
The perturbative RG scheme for the Fibonacci chain provided a natural way of explaining the multifractal properties of the spectrum and of the eigenstates. Since an analogous RG scheme is derived for the Tribonacci chains, multifractality is expected to be present. 

Since the multifractal properties of the HTC are compared to the Fibonacci chain, the definition of the Fibonacci chain is briefly reviewed here. The Fibonacci word $W^F = w^F_0 w^F_1 \cdots$ is the fixed point of the binary substitution $\rho_F : 0 \to 01, 1 \to 0$. The Fibonacci approximants are given by $W^F_N := \rho^N_F(1)$. The length of the approximants is given by the Fibonacci numbers $F_N = F_{N-1} + F_{N-2}$, where $F_0 = F_1 = 1$. The Hamiltonian for the periodic hopping Fibonacci chain reads
\begin{equation}\label{eq:fibonacci_chain}
    H_N^F = \sum_{n=0}^{F_N-1} t_{w^F_n} \ket{n+1 \bmod F_N} \bra{n} + H.c.,
\end{equation}
where the hopping parameters $t_0,t_1$ are related by $t_0/t_1 = \rho$.

\begin{figure}[tb]
    \centering
    \includegraphics[width = \linewidth]{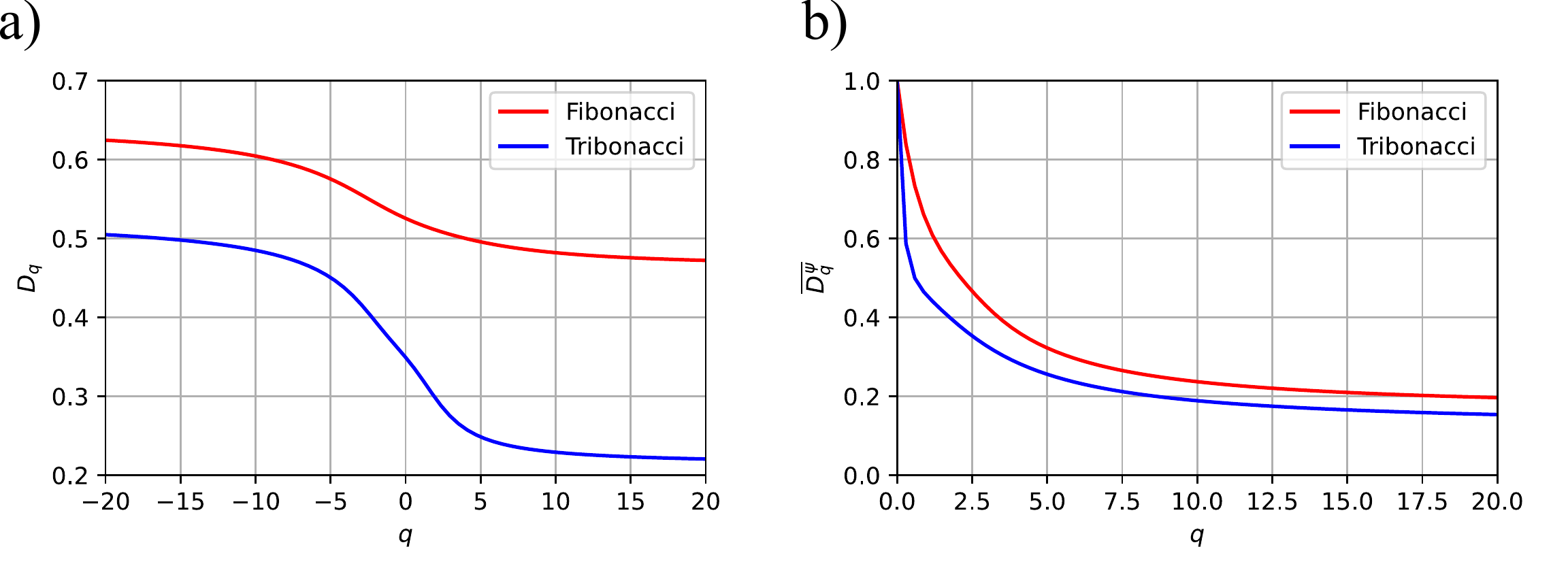}
    \caption{Multifractal properties of the Fibonacci chain Eq.~(\ref{eq:fibonacci_chain}) and the HTC, at $\rho=0.2$. (a) The multifractal dimensions of the energy spectrum for the Fibonacci chain $H^F_{19}$ and the HTC $H_{13}$. The size of the chain is chosen such that they have approximately the same maximum number of points in $K_i$. (b) The average multifractal dimensions of the eigenstates of the Fibonacci chain $H^F_{14}$ and the HTC $H_{10}$. The Size of the chains is choses such that they have approximately the same number of lattice sites.}
    \label{fig:multifractal}
\end{figure}

To study the multifractal properties of the spectrum of any Hamiltonian, we compute the multifractal dimensions $D_q$, also known as the multifractal spectrum, introduced by Halsey et al.~\cite{HALSEY_KADANOFF}. The multifractal spectrum is a family of dimensions that is continuously parameterized by $q \in \R$, where $D_0$ recovers the box-counting dimension. For the energy spectrum, the multifractal dimensions are computed as follows. First, cover the energy spectrum with a compact interval $C \subset \R$. Then, partition $C$ into intervals $K_i$ of length $l$. Let the measure $\mu(K_i)$ denote the fraction of points that lie in $K_i$. The multifractal dimensions are then given by
\begin{equation}
    D_q = \lim_{l \downarrow 0} \frac{1}{q-1} \frac{\log \sum_i \mu(K_i)^q}{\log l}.
\end{equation}
The result is shown in Fig.~\ref{fig:multifractal}(a), where the multifractal dimensions of the HTC $H_{13}$ and the Fibonacci chain $H^F_{19}$ are plotted. One can see that the HTC energy spectrum is a multifractal, since the spectrum $D_q$ is a smooth curve of $q$. Moreover, the multifractal dimensions of the HTC are strictly smaller than that of the Fibonacci chain.

For the eigenstates, the average multifractal dimension is computed. The average multifractal dimension of the eigenstates is defined as \cite{MIRLIN_MULTIFRACTAL, MACE_MULTIFRACTAL}
\begin{equation}
    \overline{D_q^\psi} = \frac{1}{q-1} \frac{\log \frac{1}{N} \sum_{i} \sum_n |\psi_i(n)|^{2q}}{\log 1/N},
\end{equation}
where the sum over $i$ ranges over all eigenstates, $N$ denotes the amount of eigenstates and $n$ ranges over the lattice sites. The numerical results for the Fibonacci chain and HTC are displayed in Fig.~\ref{fig:multifractal}(b). The average multifractal dimension of the HTC is lower than of the Fibonacci chain. This is to be expected, since $\overline{D_q^\psi}$ is related to diffusive properties of the system \cite{MIRLIN_MULTIFRACTAL}. The weakest bonds in the HTC are $\mathcal{O}(\rho^2)$, whereas in the Fibonacci they are $\mathcal{O}(\rho)$. This makes it more difficult for a particle to diffuse in the HTC than in the Fibonacci chain, which is in accordance with the fact that the average multifractal dimension for the HTC is lower than of the Fibonacci chain. Additionally, a lower average multifractal dimension indicates that the wavefunctions are more localized, which is a consequence of the weaker bonds in the HTC. In fact, the HTC is a critical chain in terms of Anderson localization, since the eigenstates are multifractal with $0<\overline{D_q^\psi}<1$ \cite{EVERS_ANDERSON}. Finally, because the OTC is approximately a direct product of HTC Hamiltonians in Eq.~(\ref{eq:tribo_renorm_result_onsite}), the multifractal properties perturbatively carry over to the OTC.

\section{Conclusion}\label{sec:conclusion}
In this work, we introduced two tight-binding chains Eqs.~(\ref{eq:infinite_tribonacci_chain_hamiltonian}) and (\ref{eq:infinite_onsite_tribonacci_chain_hamiltonian}), based on the Tribonacci substitution, which generalizes the Fibonacci chain. One of the first steps towards understanding these models are the RG Eqs.~(\ref{eq:tribo_renorm_result}) and (\ref{eq:tribo_renorm_result_onsite}), which are more accurate than those for the Fibonacci chain due to the higher orders of perturbation theory required. As shown in Section~\ref{sec:equivalence}, the two models can be regarded as equivalent at the RG fixed point. The Rauzy fractal, which is inherent to the Tribonacci word, is shown to serve as the analog of the conumbers for the HTC and OTC, since it orders the sites according to their local environment. The structure of eigenstates, when plotted on the Rauzy fractal, shows self-similar properties, which reflect the fractal nature of the eigenstates. These self-similar structures can be systematically enumerated using the RG scheme, and are exactly the LRMs within the framework proposed by R\"ontgen et al. \cite{RONTGEN_LRM}. Finally, the multifractal dimensions of both the energy spectrum and the eigenstates of the HTC have been computed, and compared to those of the Fibonacci chain. The multifractal properties are qualitatively similar to those of the Fibonacci chain, whereas the multifractal dimensions of the HTC are generally smaller than those of the Fibonacci chain. Furthermore, the HTC is shown to be a critical model in terms of Anderson localization, since the wavefunctions exhibit multifractal properties with a dimension between zero and one. 

This work opens some interesting topics for further research. First of all, it would be interesting to identify an equivalence between the HTC and another model, such as the one by Kraus and Zilberberg~\cite{KRAUS_ZILBERBERG} for the Fibonacci chain. Such an equivalence would be key to understanding the topological properties of the HTC. One could also generalize the substitution to any Pisot substitution, or consider the general $k$-bonacci substitution $0 \to 01, 1 \to 02,\dots, (k-1) \to 0$. The latter would make the generalization of the Fibonacci chain as complete as the complementary generalization in Refs.\cite{VIDAL_2001_GEN_RAUZY_TILINGS, JAGANNATHAN_RAUZY2001, VIDAL_RAUZY2002}. Yet another proposition to check is whether quasicrystals can generally be studied via their internal space, which is conumbering for the Fibonacci chain and the Rauzy fractal for the HTC, and how the RG scheme can be applied in the internal space to understand the eigenstates. Since the RG scheme originates from the self-similar structure, it could be interesting to study if self-similarity can replace translational invariance in the topological classification of quasicrystals and/or fractals. Finally, experimental realizations, such as polaritonic waveguides \cite{baboux2017measuring} and dielectric resonators \cite{reisner2022experimental} for the Fibonacci chain, can be realized to probe the electronic and multifractal properties of the HTC and OTC.



\begin{acknowledgments}
This publication is part of the project TOPCORE with Project No. OCENW.GROOT.2019.048 which is financed by the Dutch Research Council (NWO).
\end{acknowledgments}

\interlinepenalty=10000 

\bibliography{bibliography}

\interlinepenalty=1 

\appendix

\section{Connection Between the Valuation Map and the Projection Method}\label{app:rauzy}

\begin{figure}
    \includegraphics[width=\linewidth]{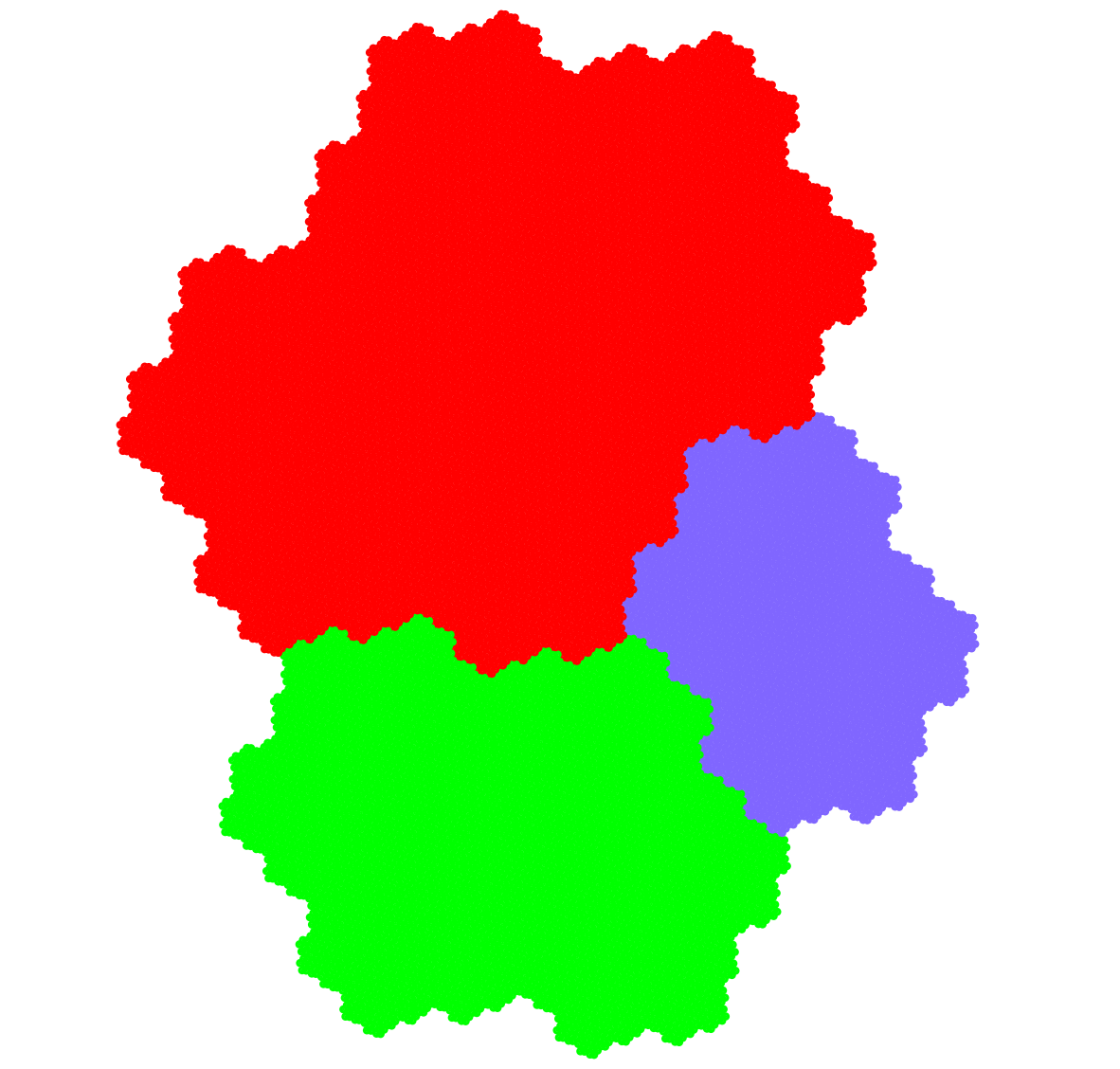}
    \caption{Rauzy fractal, using the valuation map Eq.~(\ref{eq:valuation_map}).}
    \label{fig:rauzy_valuation}
\end{figure}

\begin{figure}
    \includegraphics[width=\linewidth]{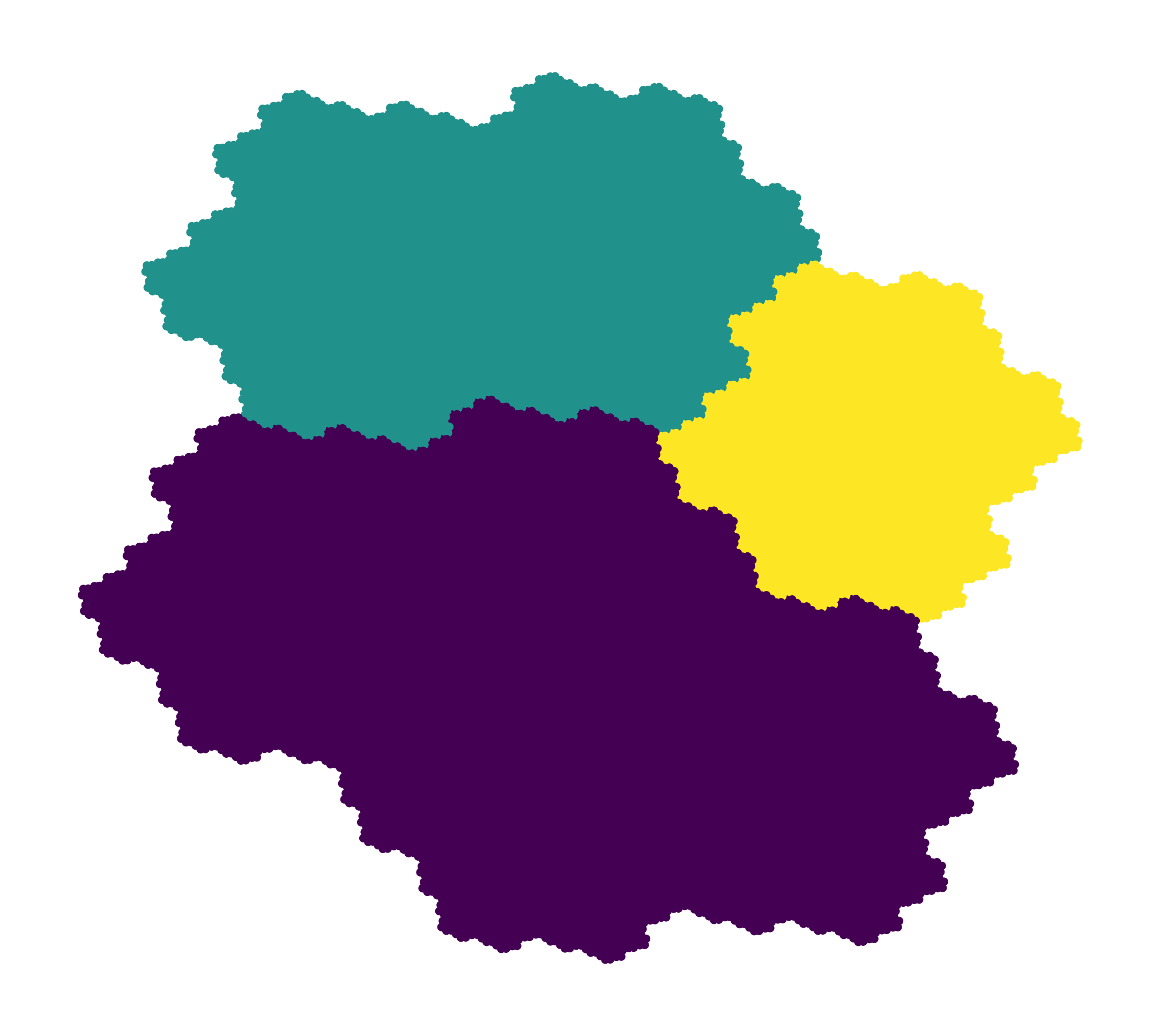}
    \caption{Rauzy fractal, using the projection method Eq.~(\ref{eq:rauzy_project_x_m}).}
    \label{fig:rauzy_projection_real}
\end{figure}

There are two main ways of generating the Rauzy fractal: projecting on the contracting eigenspace of the tribonacci substitution or using a valuation map. By using a bi-orthogonal basis of eigenvectors of the adjacency matrix of the tribonacci substitution, we can derive the affine transformation that relates the two different methods of generating the Rauzy fractal. It turns out that the valuation map generates a Rauzy fractal where the domains are scaled versions of the whole fractal, see Fig. \ref{fig:rauzy_valuation}, which is the canonical Rauzy fractal. The projection method yields a skewed version of this fractal (see Fig. \ref{fig:rauzy_projection_real}). It is worth noting that many sources, see \cite{fogg2002substitutions} (section 7.4.3) and \cite{arnoux2001pisot, arnoux2014rauzy}, claim that the canonical Rauzy fractal is obtained using the projection method, which is strictly speaking not the case.

The Rauzy fractal, as introduced in Section \ref{sec:rauzy_fractal}, was obtained by Rauzy in 1982 \cite{RAUZY} by means of a \emph{valuation map} $E$. Let $\mathcal{A} = \{0,1,2 \}$ be the alphabet for the tribonacci substitution $\rho$ and denote $\mathcal{A}^*$ the set of all finite words with letters in $\mathcal{A}$. Then the valuation map $E: \mathcal{A}^* \to \mathbb{C}$ associates a complex number to each finite word. For any $u,v \in \mathcal{A}^*$, Rauzy demanded that $E(uv) = E(u) + E(v)$ and $E(\rho(u)) = \omega E(u)$ for some constant $\omega \in \mathbb{C}$. Note that since $\rho(W) = W$, it must be true that $E(W) = 0$. Denote $|u|_i$ the number of times that $i \in \mathcal{A}$ occurs in $u$. A crucial fact is that the tribonacci substitution $\rho$ is Pisot. This implies that the adjacency matrix Eq.~(\ref{eq:tribonacci_adjacency_matrix}) has one eigenvalue $|\beta| > 1$ and all other eigenvalues have $|\lambda|<1$. The adjacency matrix $\vb{M}$ has one real eigenvalue $\beta = (1+\sqrt[3]{19 + 3\sqrt{33}}+\sqrt[3]{19-3\sqrt{33}})/3 \approx 1.8392$, the tribonacci constant, and two complex eigenvalues $\omega, \Bar{\omega}$, which are complex conjugates of each other. The corresponding normalised right eigenvectors of $\vb{M}$ are denoted by $\ket{v_\beta}, \ket{v}, \overline{\ket{v}}$. The valuation map (identical to Eq.~(\ref{eq:valuation_map})) is given by 
\begin{equation*}
    E(u) = \sum_{i \in \mathcal{A}} |u|_i v_i,
\end{equation*}
where $\bra{v^t} = (v_0,v_1,v_2)$ is the left eigenvector of $\vb{M}$, corresponding to the eigenvalue $\omega$. We emphasize that $\bra{v^t}$ is obtained via a bi-orthonormal construction. This means that the left eigenvectors need not be normalized, but the left and right eigenvectors form a bi-orthonormal set. Now, we are in shape to define the Rauzy fractal using the valuation map. Let $[W]_m$ denote the first $m$ symbols of the tribonacci word. The Rauzy fractal is the set 
\begin{equation}\label{eq:rauzy_complex}
    \mathfrak{R} : = \overline{\{ E([W]_m) \mid \forall m \in \N) \}},
\end{equation}
which is displayed in Fig. \ref{fig:rauzy_valuation}. This plot is made by taking the complex number $z = a + bi$, where $a,b\in \R$, and plotting it as a point $(a,b) \in \R^2$. This set can be partitioned in three domains $\mathfrak{R}_i$ where $i=0,1,2$ (red, green, blue respectively), that are the same as $\mathfrak{R}$ up to a factor $\beta^{-(1+i)}$, a rotation and translation. These domains are defined by 
\begin{equation}\label{eq:rauzy_complex_domains}
    \mathfrak{R}_i : = \overline{\{ E([W]_m) \mid w_m=i, \, m \in \N) \}},
\end{equation}
where $w_m$ denotes the $m$th symbol of $W$. See Ref.~\cite{SIRVENT} for more details on generating the Rauzy fractal.


\begin{figure}
    \includegraphics[width=\linewidth]{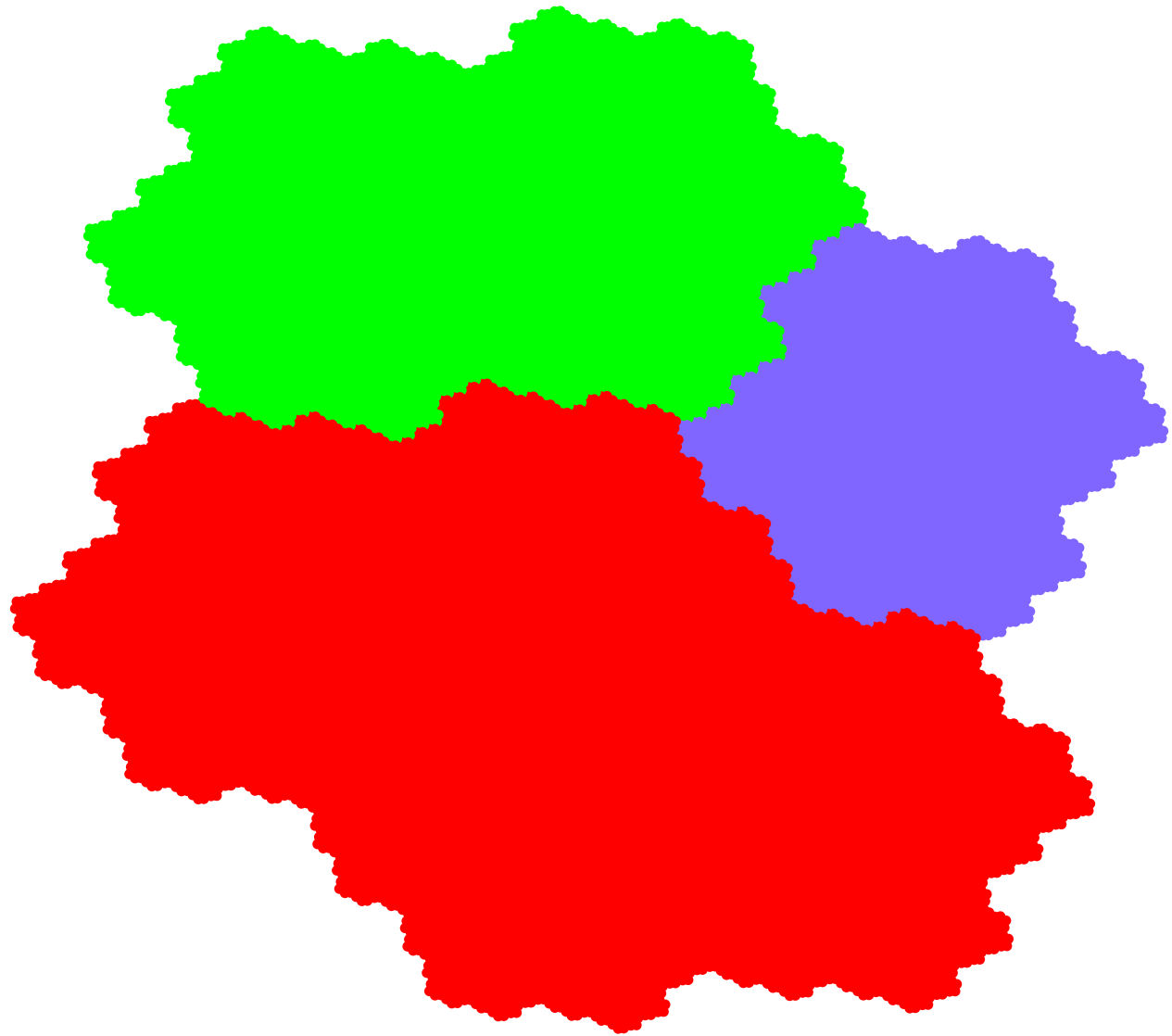}
    \caption{Result of applying affine transformation Eq.~(\ref{eq:valuation_to_project_A}) to valuation map result in Fig.~\ref{fig:rauzy_valuation}.}
    \label{fig:rauzy_projection}
\end{figure}

The second method, which we refer to as the \emph{projection method}, is the method explained in Section \ref{sec:rauzy_fractal}. Let the unstable eigenspace $E^u$ of $\vb{M}$ be the one-dimensional real space spanned by $\ket{v_\beta}$, and the stable eigenspace $E^s$ be spanned by $\{ \ket{\text{re}} , \ket{\text{im}} \}$, where $\ket{\text{re}} = \Re{\ket{v}}$ and $\ket{\text{im}} = \Im{\ket{v}}$. Equivalently, one can define $E^s = \{ c \ket{v} + \Bar{c} \overline{\ket{v}} \mid c \in \mathbb{C} \} \subset \R^3$. Denote $\pi$ the projection along $E^u$ onto $E^s$. Any $\ket{x} \in \R^3$ can be uniquely decomposed as $\ket{x} = \gamma \ket{v_\beta} + c \ket{v} + \Bar{c} \overline{\ket{v}}$ where $\gamma \in \R, c \in \mathbb{C}$. The map $\pi$ then acts as $\pi \ket{x} = c \ket{v} + \Bar{c} \overline{\ket{v}}$. Denoting the $m$th point in the staircase as $\ket{x_m} = \sum_{i=0}^m \ket{e_{w_i}}$, the projection method yields the Rauzy fractal 
\begin{equation*}
    R = \overline{\{ \pi \ket{x_m} \mid m \in \N \}},
\end{equation*}
which is displayed in Fig. \ref{fig:rauzy_projection_real}. One immediate connection between the valuation and the projection method, is the fact that the valuation map can be written as
\begin{equation*}
    E([W]_m) = \braket{v^t}{x_m} = c,
\end{equation*}
where we used $\ket{x_m} = \gamma \ket{v_\beta} + c \ket{v} + \Bar{c} \overline{\ket{v}}$ and the fact that $\{ \bra{\smash{v_\beta^t}}, \bra{v^t}, \overline{\bra{v^t}} \}$ and $\{ \ket{v_\beta}, \ket{v}, \overline{\ket{v}} \}$ form a bi-orthogonal system.

The question now is: how are the sets in Figs \ref{fig:rauzy_valuation} and \ref{fig:rauzy_projection_real} related? We will derive an affine transformation that maps the former to the latter as points in $\R^2$. We first need a choice of basis to represent the points of $\mathfrak{R}$ and $R$ in $\R^2$. For the complex numbers in $\mathfrak{R}$, we have the canonical representation $z = a+bi \mapsto (a,b) \in \R^2$. In associating $E^s$ with $\R^2$, we have some freedom. By applying a Gram-Schmidt procedure to the real and imaginary parts of $\ket{v}$, we can define 
\begin{equation} \label{eq:rauzy_e1e2}
\begin{split}
\ket{e_1} &= \ket{\text{re}} / \norm{\ket{\text{re}}}, \\
\ket{e_2} &= \ket{\text{res}} / \norm{\ket{\text{res}}},
\end{split}
\end{equation}
where $\ket{\text{res}} = \ket{\text{im}} - \braket{e_1}{\text{im}} \ket{e_1}$. Then we can represent any $\ket{x} = a' \ket{e_1} + b' \ket{e_2} \in E^s \subset \R^3$ as $(a',b') \in \R^2$.

We can now answer the question: for a given $\ket{x_m}$, how are $(a,b)$ and $(a',b')$ related? For a fixed $m \in \N$, we have seen that $E([W]_m) = c = a + bi$ and $\pi \ket{x_m} = c \ket{v} +\Bar{c} \overline{\ket{v}} = a' \ket{e_1} + b' \ket{e_2}$. The answer is the matrix $A$ that solves
\begin{equation}\label{eq:valuation_to_project_map}
\begin{pmatrix}
a' \\
b' 
\end{pmatrix} = 
A 
\begin{pmatrix}
a \\
b 
\end{pmatrix}.
\end{equation}
By noting that 
\begin{equation}\label{eq:rauzy_deriv_1}
    \ket{x} = c \ket{v} +\Bar{c} \overline{\ket{v}} = 2 a \ket{\text{re}} - 2 b \ket{\text{im}},
\end{equation}
and inverting Eqs.~(\ref{eq:rauzy_e1e2}) to obtain
\begin{equation}\label{eq:rauzy_deriv_2}
\begin{split}
    \ket{\text{re}} &= \norm{\ket{\text{re}}} \ket{e_1}, \\
    \ket{\text{im}} &= \norm{\ket{\text{res}}} \ket{e_2} + \braket{e_1}{\text{im}} \ket{e_1},
\end{split}
\end{equation}
we can plug Eqs.~(\ref{eq:rauzy_deriv_2}) into Eq.~(\ref{eq:rauzy_deriv_1}) to obtain
\begin{equation*}
    \ket{x} = (2 a \norm{\ket{\text{re}}} -2b \braket{e_1}{\text{im}} )\ket{e_1} - 2 b \norm{\ket{\text{res}}} \ket{e_2},
\end{equation*}
from which we can read off the matrix $A$ as being given by
\begin{equation}\label{eq:valuation_to_project_A}
    A = 
    \begin{pmatrix}
2 \norm{\ket{\text{re}}} & -2 \braket{e_1}{\text{im}} \\
0 & -2 \norm{\ket{\text{res}}} 
\end{pmatrix}.
\end{equation}
To demonstrate the correctness of Eq.~(\ref{eq:valuation_to_project_map}), the map in Eq.~(\ref{eq:valuation_to_project_A}) is applied to $\mathfrak{R}$ in Fig. \ref{fig:rauzy_valuation}, which yields the result in Fig. \ref{fig:rauzy_projection}. It is of no surprise that Figs. \ref{fig:rauzy_projection} and \ref{fig:rauzy_projection_real} are identical, since we have just derived the mathematical correspondence between the two. 

\section{Brillouin-Wigner Perturbation Theory Computations}\label{App:BWPT}

This section explains how perturbation theory can be applied to $H_N$ to compute the values of $z_i$ and the $p,q$-exponents of the blocks in Eqs.~(\ref{eq:tribo_renorm_result}) and (\ref{eq:tribo_renorm_result_onsite}). Canonical Rayleigh-Schr\"odinger time-independent perturbation theory cannot be applied here because of the degeneracy of the $H_{0,N}$ energy levels. This problem also arose in the case of the Fibonacci chain, for which Brillouin-Wigner perturbation theory was used. The perturbation theory frameworks will be introduced, after which we apply it to the Tribonacci chain. We start by writing the Hamiltonian as
\begin{equation*}
    H = H_0 + H_1,
\end{equation*}
analogous to Eq.~(\ref{eq:H_N_split}).
Denote $Q$ as the projection operator on the eigenspace of $E_0$, which is spanned by all $\ket{\psi_0}$ such that $H_0 \ket{\psi_0} = E_0 \ket{\psi_0}$. The goal is to derive some effective Hamiltonian $H_\text{eff}$ that agrees with $H$ on $Q$, for each of the five unperturbed energy levels $E_0 = \pm t_2, \pm t_1, 0$.

We start by fixing some $\ket{\psi}, E$ that satisfy $H \ket{\psi} = E \ket{\psi}$. Additionally, define $P$ as
\begin{equation*}
     P = \operatorname{Id} - \, Q,
\end{equation*}
i.e. the projection orthogonal to $Q$. Rewriting the Schr\"odinger equation we obtain
\begin{equation}\label{eq:bwpt1}
    (E-H_0) \ket{\psi} = H_1 \ket{\psi}.
\end{equation}
Using Eq.~(\ref{eq:bwpt1}), one can check that 
\begin{equation}\label{eq:bwpt_P}
    P \ket{\psi} = P \frac{1}{E-H_0} H_1 \ket{\psi} 
\end{equation}
holds. We denote the inverse of an operator $\mathcal{O}$ as $\mathcal{O}^{-1} = \frac{1}{\mathcal{O}}$ for clarity in future computations. By noting that $H_0 P = P H_0$ such that 
\begin{equation}\label{eq:bwpt2}
    (E-H_0) P \ket{\psi} = P H_1 \ket{\psi},
\end{equation}
and using $P^2 = P$ and Eqs.~(\ref{eq:bwpt1}) and (\ref{eq:bwpt2}), one can check that the following equalities:
\begin{multline}\label{eq:BWPT_P_trick}
    P \ket{\psi} = P \frac{1}{E-H_0} H_1 \ket{\psi} =  \frac{1}{E-H_0} P H_1 \ket{\psi} \\ = P \frac{1}{E-H_0} P H_1 \ket{\psi}.
\end{multline}
Note that $\frac{1}{E-H_0}$ is ill-defined (division by zero) precisely on the kernel of $P$. Hence, every time one needs to find an expression for $P \frac{1}{E-H_0}$, the well-defined expression for $\frac{1}{E-H_0} P$ can be used instead. Using $P$ and $Q$ we write $\ket{\psi}$ as
\begin{equation}\label{eq:bwpt_iter}
    \ket{\psi} = (Q+P) \ket{\psi} = Q \ket{\psi} + P \frac{1}{E-H_0} H_1 \ket{\psi}.
\end{equation}
Realising that Eq.~(\ref{eq:bwpt_iter}) is a self-consistent equation for $\ket{\psi}$, one can sum all the terms that arise from iterating that equation to get
\begin{equation}\label{eq:bwpt_sumketpsi}
    \ket{\psi} = \left[ \sum^\infty_{n=0} \brac{P \frac{1}{E-H_0} H_1}^n \right] Q \ket{\psi}.
\end{equation}
To obtain an expression for $H_\text{eff}$, one needs to multiply Eq.~(\ref{eq:bwpt_sumketpsi}) with $Q H$ which yields
\begin{equation*}
    E Q \ket{\psi} =\underbrace{ \left[ Q H_0 + Q H_1 \sum^\infty_{n=0} \brac{P \frac{1}{E-H_0} H_1}^n \right]}_{H_\text{eff}} Q \ket{\psi}. 
\end{equation*}
Using $Q^2 = Q$, the expression for $H_\text{eff}$ can be written as
\begin{equation}\label{eq:BWPT_Heff}
    H_\text{eff} = Q H_0 Q+ Q H_1 \left[ \sum^\infty_{n=0} \brac{P \frac{1}{E-H_0} H_1}^n \right] Q.
\end{equation}
Note that during the whole derivation, there was no need to pick a certain unperturbed energy $E_0$, which is strictly needed to define the projectors $P,Q$. Suppose one now chooses some $E_0$, one might wonder what value to insert for $E$, which is the unknown energy of the full system. It turns out that the energy $E_0$ of the unperturbed system can be used, as long as $H_\text{eff}$ is used at the lowest non-vanishing order in $H_1$ for that particular computation. 

For further reference, we start by writing down all the orders of Eq.~(\ref{eq:BWPT_Heff}) that we need in this appendix: 
\begin{align*}
    H_\text{eff}^{(0)} =& Q H_0 Q, \\
    H_\text{eff}^{(1)} =& H_\text{eff}^{(0)} + Q H_1 Q, \\
    H_\text{eff}^{(2)} =& H_\text{eff}^{(1)} + Q H_1 P \frac{1}{E-H_0} H_1 Q, \\
    H_\text{eff}^{(3)} =& H_\text{eff}^{(2)} + Q H_1 P \frac{1}{E-H_0} H_1 P \frac{1}{E-H_0} H_1 Q, \\
    \vdots& \\
    H_\text{eff}^{(n)} =& H_\text{eff}^{(n-1)} +QH_1 \brac{P \frac{1}{E-H_0}H_1}^{n-1}Q. 
\end{align*}

Before showing the computations for all renormalized couplings, let us work out one simple example of how the RG scheme is applied to a small Tribonacci approximant. Consider the HTC chain $H_3$ with $T_3=7$ sites with PBC, which is displayed in Fig.~\ref{fig:tribo_renorm_chains}. In the unperturbed Hamiltonian
\begin{equation}
    H_{0,3} = t_1 \ket{2} \bra{3} + t_2 \ket{4} \bra{5}+ t_1 \ket{6} \bra{7} +H.c.,
\end{equation}
the type-1 molecules are the sites of the Hamiltonian coupled by a $t_1$ bond, on which the eigenstates read
\begin{align}
    H^\text{m,1}_1 \ket{\pm}_1 = \pm t_1, \quad \ket{\pm}_1 = \frac{\ket{2} \pm \ket{3}}{\sqrt{2}}, \\
    H^\text{m,1}_2 \ket{\pm}_3 = \pm t_1, \quad \ket{\pm}_3 = \frac{\ket{6} \pm \ket{7}}{\sqrt{2}}.
\end{align}
The new lattice sites $\ket{+}_1, \ket{+}_3$ constitute the type-1 molecular bonding chain and $\ket{-}_1, \ket{-}_3$ the corresponding anti-bonding chain. The renormalized hopping parameters are
\begin{equation}\label{eq:renorm_couplings_example}
    t_0' = \bra{+}_1 H_{020} \ket{+}_3, \quad t_1' = \bra{+}_3 H_{00} \ket{+}_1,
\end{equation}
where $H_{020} = t_0(\ket{3} \bra{4} + \ket{5} \bra{6}) + t_2 \ket{4} \bra{5} +H.c.$ and $H_{00} = t_0(\ket{7} \bra{1} + \ket{1} \bra{2}) + H.c.$ and the matrix elements are evaluated using perturbation theory. In this case, the projector $Q$ reads
\begin{equation*}
    Q_\pm = \ket{\pm}_1 \bra{\pm}_1 + \ket{\pm}_3 \bra{\pm}_3, 
\end{equation*}
which projects out all states except the unperturbed eigenstates with $E_0 = \pm t_1$. How to evaluate the matrix elements in Eq.~(\ref{eq:renorm_couplings_example}) using perturbation theory is explained in the next section.

\begin{figure*}[tb]
    \includegraphics[width = \linewidth]{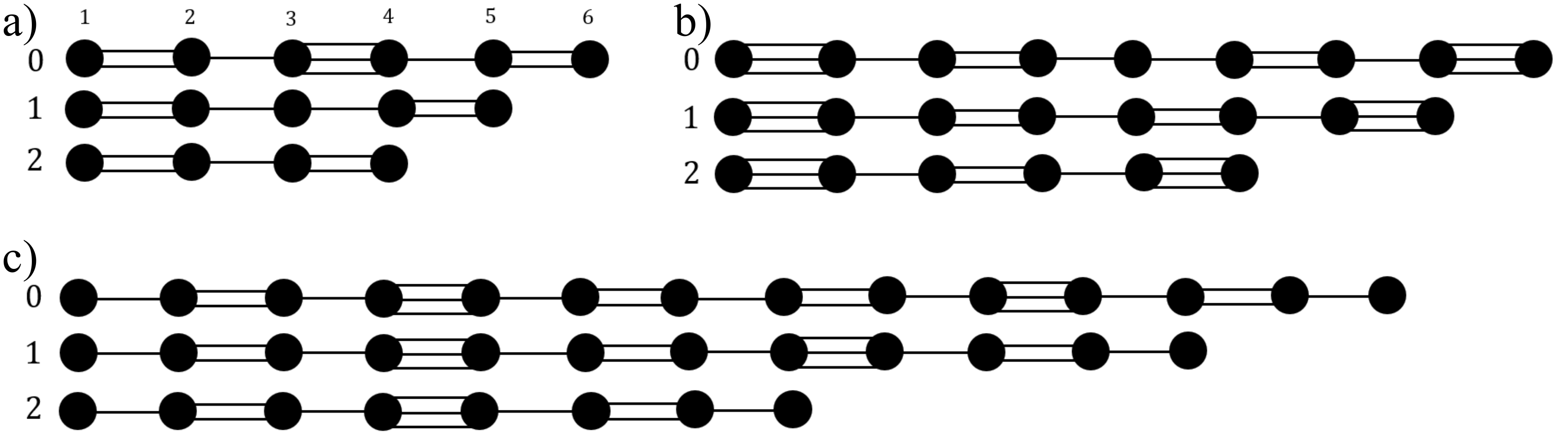}
    \caption{The different chains occurring in the HTC. The black dots denote the lattice sites, a single/double/triple line denotes a $t_0$/$t_1$/$t_2$ bond. a) the molecular-1 chains, b) molecular-2 chains and c) atomic chains. The vertical numbers $0,1,2$ denote the letter to which that chain renormalizes, the horizontal numbers $1,2,\dots$ denote the lattice site labeling in the perturbative calculations in Section~\ref{App:BWPT_hopping}.}
    \label{fig:atom_molec_chains}
\end{figure*}

\subsection{Hopping Model}\label{App:BWPT_hopping}


The results of the perturbative calculations for the renormalized couplings $t'_i$ are summarized in Table~\ref{tab:tribo_renorm_couplings_HTC}. In this table, the function $c(p,q)$ is the leading order in $\rho$ in $t'_i = c(p,q) t_i$. The last two columns give the new $\rho^{p_i} = |t_0'/t_1'|$ and $\rho^{q_i} = |t_1'/t_2'|$.

We can now explain how Eq.~(\ref{eq:tribo_renorm_result}) is obtained. Let $t_0=0$, take one of the five unperturbed energies $E_0 = 0,\pm t_1, \pm t_2$, and consider the chain formed by the unperturbed eigenstates with that energy. Firstly, the $p_i$ and $q_i$ corresponding to that chain are exactly the exponents in the block $z_i H^{(p_i,q_i)}_{N-k} + E_0$. The value $k$ corresponding to $E_0$ can be read off from the last column of Table~\ref{tab:tribo_renorm_words}, and is given by $k = 2,3,4$ for $E_0 = \pm t_1, \pm t_2, 0$ for the HTC, respectively. The value of $z_i$ is the ratio between the chain under consideration and $H^{(p_i,q_i)}_{N-k}$. Note that if all bonds in $H^{(p_i,q_i)}_{N-k}$ are divided by $t_2$, the spectrum is bounded and of size $\mathcal{O}(1)$. So in order to match the energy scale of the chain under consideration and $H^{(p_i,q_i)}_{N-k}$, we divide all renormalized couplings in the chain by $t_2'$ and all couplings in $H^{(p_i,q_i)}_{N-k}$ by $t_2$. This is equivalent to multiplying $H^{(p_i,q_i)}_{N-k}$ by $t'_2/t_2 = c_2(p,q) = z_i$, which means that $z_i H^{(p_i,q_i)}_{N-k}$ is now an approximation of the HTC with renormalized couplings $t'_i$. Finally, the result is shifted by the value of $E_0$, since the new lattice sites of each chain have the on-site energy $E_0$. This procedure is carried out for each of the five $E_0$ values, thereby obtaining five blocks in Eq.~(\ref{eq:tribo_renorm_result}). 

The remainder of this section is devoted to computing the $t'_i$ values in Table~\ref{tab:tribo_renorm_couplings_HTC}. All calculations are done for the HTC $H^{(p,q)}_N$, where $p,q>0$ are assumed to be integers. It turns out that the order of perturbation theory needed is equal to the amount of $t_0$ in the chain that is considered in Fig.~\ref{fig:atom_molec_chains}. This means that in the worst case, which is the top chain in Fig.~\ref{fig:atom_molec_chains}(c), one needs seven orders of perturbation theory. We show all computational details for the type-1 molecular chain. For the other chains, the computational steps are identical, but we give only the operators $H_1,\frac{1}{E-H_0}P$ and eigenstates, and skip the computational steps of repeatedly applying these operators. This is enough, because if the operators and eigenstates are known, the computation can be carried out using a computer algebra program such as Mathematica.

\subsubsection{Type-1 Molecules}
\paragraph{Computation of $t'_0$}
Consider the top chain in Fig.~\ref{fig:atom_molec_chains}(a), for which the Hamiltonian $H=H_0+H_1$ reads
\begin{align*}
    H_0 &= t_1\ket{1}\bra{2} + t_2\ket{3}\bra{4} + t_1\ket{5}\bra{6} +H.c., \\
    H_1 &= t_0 \ket{2}\bra{3} + t_0 \ket{4}\bra{5} +H.c.,
\end{align*}
and the six eigenstates of $H_0$ read
\begin{align*}
    \ket{\pm}_1 &=  \brac{\ket{1} \pm \ket{2}}/\sqrt{2}, \qquad E_0 = \pm t_1,\\
    \ket{\pm}_2 &= \brac{\ket{3} \pm \ket{4}} / \sqrt{2}, \qquad E_0 = \pm t_2,\\
    \ket{\pm}_3 &= \brac{\ket{5} \pm \ket{6}} / \sqrt{2}, \qquad E_0 = \pm t_1,
\end{align*}
from which we can read off
\begin{equation*}
    Q = \ket{\pm}_1 \bra{\pm}_1 + \ket{\pm}_3 \bra{\pm}_3.
\end{equation*}
The perturbation theory gives
\begin{align*}
    \bra{\pm}_1 H_\text{eff}^{(0)} \ket{\pm}_3 =& \bra{\pm}_1 H_0 \ket{\pm}_3 =\pm t_1 \bra{\pm}_1  \ket{\pm}_3=0, \\
    \bra{\pm}_1 H_\text{eff}^{(1)} \ket{\pm}_3 =& \bra{\pm}_1 H_1 \ket{\pm}_3 = \pm \frac{t_0}{\sqrt{2}}\bra{3} \ket{\pm}_3 = 0, \\
    \bra{\pm}_1 H_\text{eff}^{(2)} \ket{\pm}_3 =& \bra{\pm}_1 H_1 P \frac{1}{\pm t_1 - H_0} H_1 \ket{\pm}_3 \\
    =& \pm \frac{t_0^2}{2} \bra{3} P \frac{1}{\pm t_1 - H_0} \ket{4}.
\end{align*}
We can write $\frac{1}{\pm t_1 - H_0} P$ as
\begin{multline*}
    \frac{1}{\pm t_1 - H_0}P = \frac{1}{\pm 2 t_1} \ket{\mp}_1 \bra{\mp}_1 + \frac{1}{\pm t_1 \mp t_2} \ket{\pm}_2 \bra{\pm}_2 \\ + \frac{1}{\pm t_1 \pm t_2} \ket{\mp}_2 \bra{\mp}_2 +\frac{1}{\pm 2 t_1} \ket{\mp}_3 \bra{\mp}_3,
\end{multline*}
and by using $\ket{4} =\pm \frac{\ket{\pm}_2 - \ket{\mp}_2}{\sqrt{2}}$ we can proceed as
\begin{align*}
    \bra{\pm}_1 & H_\text{eff}^{(2)} \ket{\pm}_3 = \frac{t_0^2}{2 \sqrt{2}} \bra{3}  \frac{1}{\pm t_1 - H_0}P (\ket{\pm}_2 - \ket{\mp}_2) \\
    =& \pm \frac{t_0^2}{2 \sqrt{2}} \bra{3} \brac{ \frac{1}{ t_1 - t_2}\ket{\pm}_2 - \frac{1}{ t_1 + t_2}\ket{\mp}_2} \\
    =& \mp \frac{t_0^2}{2} \frac{t_2}{t_2^2-t_1^2} = t'_0.
\end{align*}
This can be approximated as
\begin{align*}
    t'_0 = \mp \frac{t_0^2}{2 t_2} \brac{1 + \mathcal{O}\brac{\frac{t_1^2}{t_2^2}}} \approx \mp \frac{t_0^2}{2 t_2} = \mp \frac{\rho^{p+q}}{2} t_0,
\end{align*}
such that $c_0(p,q) = \rho^{p+q}/2$ can be read off.

\paragraph{Computation of $t'_1$}
Consider the middle chain in Fig.~\ref{fig:atom_molec_chains}(a), for which the Hamiltonian $H=H_0+H_1$ reads
\begin{align*}
    H_0 &= t_1\ket{1}\bra{2} + t_1\ket{4}\bra{5} +H.c., \\
    H_1 &= t_0 \ket{2}\bra{3} + t_0 \ket{3}\bra{4} +H.c.,
\end{align*}
and the five eigenstates of $H_0$ read
\begin{align*}
    \ket{\pm}_1 &=  \brac{\ket{1} \pm \ket{2}}/\sqrt{2}, & \qquad E_0 &= \pm t_1,\\
    \ket{\psi}_2 &= \ket{3}, & \qquad E_0 &= 0,\\
    \ket{\pm}_3 &= \brac{\ket{4} \pm \ket{5}} / \sqrt{2}, & \qquad E_0 &= \pm t_1,
\end{align*}
from which we can read off
\begin{equation*}
    Q = \ket{\pm}_1 \bra{\pm}_1 + \ket{\pm}_3 \bra{\pm}_3.
\end{equation*}
Since the chain consists of two $t_0$ bonds, we need two orders of perturbation theory for a nonzero result. For that we need
\begin{multline*}
    \frac{1}{\pm t_1 - H_0}P = \frac{1}{\pm 2 t_1} \ket{\mp}_1 \bra{\mp}_1 \\ + \frac{1}{\pm t_1} \ket{\psi}_2 \bra{\psi}_2  +\frac{1}{\pm 2 t_1} \ket{\mp}_3 \bra{\mp}_3,
\end{multline*}
which can be used to compute 
\begin{multline*}
    t'_1 = \bra{\pm}_1 H^{(2)}_\text{eff} \ket{\pm}_3 = \bra{\pm}_1 H_1 \frac{1}{\pm t_1 - H_0} P H_1 \ket{\pm}_3 \\ = \pm \frac{t_0^2}{2} \bra{3} \frac{1}{\pm t_1 - H_0} P \ket{3} = \frac{t_0^2}{2t_1}.
\end{multline*}
Finally, we can compute the ratio
\begin{equation*}
    |t'_0/t'_1|= \frac{t_0^2 t_2}{2(t_2^2-t_1^2)} \frac{2t_1}{t_0^2} = \frac{t_1}{t_2} \frac{1}{1-\frac{t_1^2}{t_2^2}} \approx \rho^q,
\end{equation*}
to leading order in $\rho$.

\paragraph{Computation of $t'_2$}

Consider the bottom chain in Fig.~\ref{fig:atom_molec_chains}(a), for which the Hamiltonian $H=H_0+H_1$ reads
\begin{align*}
    H_0 &= t_1\ket{1}\bra{2} + t_1\ket{3}\bra{4}) +H.c., \\
    H_1 &= t_0 \ket{2}\bra{3} + H.c.,
\end{align*}
and the four eigenstates of $H_0$ read
\begin{align*}
    \ket{\pm}_1 &=  \brac{\ket{1} \pm \ket{2}}/\sqrt{2}, & \qquad E_0 &= \pm t_1,\\
    \ket{\pm}_2 &= \brac{\ket{4} \pm \ket{5}} / \sqrt{2}, & \qquad E_0 &= \pm t_1,
\end{align*}
from which we can read off
\begin{equation*}
    Q = \ket{\pm}_1 \bra{\pm}_1 + \ket{\pm}_2 \bra{\pm}_2.
\end{equation*}
Since the chain consists of one $t_0$ bond, we need one order of perturbation theory for a nonzero result. Hence we can directly compute
\begin{multline*}
    t'_2 = \bra{\pm}_1 H^{(1)}_\text{eff} \ket{\pm}_2 = \bra{\pm}_1 H_1  \ket{\pm}_2 \\ =  \frac{t_0}{\sqrt{2}} \bra{\pm}_1  \ket{2} = \pm \frac{t_0}{2}.
\end{multline*}
Finally, we can compute the ratio
\begin{equation*}
    |t'_1/t'_2|= \frac{t_0^2}{2t_1} \frac{2}{t_0} = \frac{t_0}{t_1} = \rho^p.
\end{equation*}

\begin{table*}[htb]
\caption{\label{tab:tribo_renorm_couplings_HTC}Renormalized couplings for the HTC.}
\centering
\begin{tabular}{||c | c c | c c | c c | c c ||} 
 \hline
 $E_0$ & $t'_0$ &  $c_0(p,q)$ & $t'_1$ &  $c_1(p,q)$ & $t'_2$ &  $c_2(p,q)=z_i$  & $ |t'_0/t'_1|$ & $|t'_1/t'_2|$\\ [0.5ex] 
 \hline\hline
 0 & $\frac{t_0^7}{t_1^4 t_2^2}$ & $\rho^{6p+2q}$ & $-\frac{t_0^6}{t_1^3 t_2^2}$ & $\rho^{6p+2q}$ & $-\frac{t_0^4}{t_1^2 t_2}$ & $\rho^{4p+2q}$ & $\rho^{p}$ & $\rho^{2p+q}$ \\ 
 \hline
 $\pm t_1$ & $\mp \frac{t_0^2 t_2}{2(t_2^2-t_1^2)}$ & $\rho^{p+q}/2$ & $\frac{t_0^2}{2t_1}$ & $\rho^{2p}/2$ & $\pm \frac{t_0}{2}$ & $\rho^{p+q}/2$ & $\rho^{q}$ & $\rho^{p}$ \\
 \hline
 $\pm t_2$ & $\frac{t_0^4 t_1^2}{2t_2(t_1^2-t_2^2)^2}$ & $\rho^{3p+5q}/2$ & $\pm \frac{t_0^3 t_1^2}{2(t_1^2-t_2^2)^2}$ & $\rho^{3p+4q}/2$ & $\pm \frac{t_0^2 t_1}{2(t_2^2-t_1^2)}$ & $\rho^{2p+3q}/2$ & $\rho^{p+q}$ & $\rho^{p+2q}$ \\
 \hline
\end{tabular}
\end{table*}

\subsubsection{Type-2 Molecules}

\paragraph{Computation of $t'_0$}
Consider the top chain in Fig.~\ref{fig:atom_molec_chains}(b), for which the Hamiltonian $H=H_0+H_1$ reads
\begin{align*}
    H_0 &= t_2\ket{1}\bra{2} + t_1\ket{3}\bra{4} + t_1\ket{6}\bra{7} + t_2\ket{8}\bra{9} +H.c., \\
    H_1 &= t_0 \brac{\ket{2}\bra{3} +\ket{4}\bra{5} +\ket{5}\bra{6} +\ket{7}\bra{8}} + H.c.,
\end{align*}
and the nine eigenstates of $H_0$ read
\begin{align*}
    \ket{\pm}_1 &=  \brac{\ket{1} \pm \ket{2}}/\sqrt{2}, & \qquad E_0 &= \pm t_2,\\
    \ket{\pm}_2 &= \brac{\ket{3} \pm \ket{4}} / \sqrt{2}, & \qquad E_0 &= \pm t_1, \\ 
    \ket{\psi}_3 &=  \ket{5}, & \qquad E_0 &= 0,\\
    \ket{\pm}_4 &=  \brac{\ket{6} \pm \ket{7}}/\sqrt{2}, & \qquad E_0 &= \pm t_1,\\
    \ket{\pm}_5 &= \brac{\ket{8} \pm \ket{9}} / \sqrt{2}, & \qquad E_0 &= \pm t_2, 
\end{align*}
from which we can read off
\begin{equation*}
    Q = \ket{\pm}_1 \bra{\pm}_1 + \ket{\pm}_5 \bra{\pm}_5,
\end{equation*}
and 
\begin{multline*}
    \frac{1}{\pm t_2 - H_0} P = \frac{1}{\pm 2t_2} \ket{\mp}_1 \bra{\mp}_1 \\ + \frac{1}{\pm t_2 \mp t_1} \ket{\pm}_2 \bra{\pm}_2 + \frac{1}{\pm t_2 \pm t_1} \ket{\mp}_2 \bra{\mp}_2 \\ + \frac{1}{\pm t_2} \ket{\psi}_3 \bra{\psi}_3  + \frac{1}{\pm t_2 \mp t_1} \ket{\pm}_4 \bra{\pm}_4 \\ + \frac{1}{\pm t_2 \pm t_1} \ket{\mp}_4 \bra{\mp}_4 + \frac{1}{\pm 2t_2} \ket{\mp}_5 \bra{\mp}_5.
\end{multline*}
Now we can compute
\begin{multline*}
    t'_0 = \bra{\pm}_1 H^{(4)}_\text{eff} \ket{\pm}_5 \\ = \bra{\pm}_1 H_1 \brac{P \frac{1}{\pm t_2 -H_0}H_1}^{3} \ket{\pm}_5 = \frac{t_0^4 t_1^2}{2 t_2(t_1^2 - t_2^2)^2}.
\end{multline*}
To leading order in $\rho$ this can be expanded as
\begin{equation*}
    t'_0 = \frac{t_0^3 t_1^2}{2 t_2(t_1^2 - t_2^2)^2} t_0 \approx \frac{t_0^3 t_1^2}{2 t_2^5} t_0 = \frac{\rho^{3p+5q}}{2} t_0,
\end{equation*}
from which we can read off $c_0(p,q) = \rho^{3p+5q}/2$.

\paragraph{Computation of $t'_1$}
Consider the middle chain in Fig.~\ref{fig:atom_molec_chains}(b), for which the Hamiltonian $H=H_0+H_1$ reads
\begin{align*}
    H_0 &= t_2\ket{1}\bra{2} + t_1\ket{3}\bra{4} + t_1\ket{5}\bra{6} + t_2\ket{7}\bra{8} +H.c., \\
    H_1 &= t_0 \brac{\ket{2}\bra{3} +\ket{4}\bra{5} +\ket{6}\bra{7}} + H.c.,
\end{align*}
and the eight eigenstates of $H_0$ read
\begin{align*}
    \ket{\pm}_1 &=  \brac{\ket{1} \pm \ket{2}}/\sqrt{2}, & \qquad E_0 &= \pm t_2,\\
    \ket{\pm}_2 &= \brac{\ket{3} \pm \ket{4}} / \sqrt{2}, & \qquad E_0 &= \pm t_1, \\ 
    \ket{\pm}_3 &=  \brac{\ket{5} \pm \ket{6}}/\sqrt{2}, & \qquad E_0 &= \pm t_1,\\
    \ket{\pm}_4 &= \brac{\ket{7} \pm \ket{8}} / \sqrt{2}, & \qquad E_0 &= \pm t_2, 
\end{align*}
from which we can read off
\begin{equation*}
    Q = \ket{\pm}_1 \bra{\pm}_1 + \ket{\pm}_4 \bra{\pm}_4,
\end{equation*}
and 
\begin{multline*}
    \frac{1}{\pm t_2 - H_0} P = \frac{1}{\pm 2t_2} \ket{\mp}_1 \bra{\mp}_1 \\ + \frac{1}{\pm t_2 \mp t_1} \ket{\pm}_2 \bra{\pm}_2 + \frac{1}{\pm t_2 \pm t_1} \ket{\mp}_2 \bra{\mp}_2 \\ + \frac{1}{\pm t_2 \mp t_1} \ket{\pm}_3 \bra{\pm}_3 \\ + \frac{1}{\pm t_2 \pm t_1} \ket{\mp}_3 \bra{\mp}_3 + \frac{1}{\pm 2t_2} \ket{\mp}_4 \bra{\mp}_4.
\end{multline*}
Now we can compute
\begin{multline*}
    t'_1 = \bra{\pm}_1 H^{(3)}_\text{eff} \ket{\pm}_4 \\ = \bra{\pm}_1 H_1 \brac{P \frac{1}{\pm t_2-H_0}H_1}^{2} \ket{\pm}_4 = \pm \frac{t_0^3 t_1^2}{2(t_1^2-t_2^2)^2}.
\end{multline*}
To leading order in $\rho$ this can be expanded as
\begin{equation*}
    t'_1 = \pm \frac{t_0^3 t_1}{2(t_1^2-t_2^2)^2} t_1 \approx \frac{t_0^3 t_1}{2 t_2^4} \pm t_1 = \pm \frac{\rho^{3p+4q}}{2} t_1,
\end{equation*}
from which we can read off $c_1(p,q) = \rho^{3p+4q}/2$. Finally, we can compute the ratio
\begin{equation*}
    |t'_0/t'_1| = \frac{t_0^4 t_1^2}{2 t_2(t_1^2 - t_2^2)^2} \frac{2(t_1^2-t_2^2)^2}{t_0^3 t_1^2} = \frac{t_0}{t_2} = \rho^{p+q}.
\end{equation*}

\paragraph{Computation of $t'_2$}
Consider the bottom chain in Fig.~\ref{fig:atom_molec_chains}(b), for which the Hamiltonian $H=H_0+H_1$ reads
\begin{align*}
    H_0 &= t_2\ket{1}\bra{2} + t_1\ket{3}\bra{4} + t_2\ket{5}\bra{6} + H.c., \\
    H_1 &= t_0 \ket{2}\bra{3} + t_0 \ket{4}\bra{5} + H.c.,
\end{align*}
and the six eigenstates of $H_0$ read
\begin{align*}
    \ket{\pm}_1 &=  \brac{\ket{1} \pm \ket{2}}/\sqrt{2}, & \qquad E_0 &= \pm t_2,\\
    \ket{\pm}_2 &= \brac{\ket{3} \pm \ket{4}} / \sqrt{2}, & \qquad E_0 &= \pm t_1, \\ 
    \ket{\pm}_3 &=  \brac{\ket{5} \pm \ket{6}}/\sqrt{2}, & \qquad E_0 &= \pm t_2,\\
\end{align*}
from which we can read off
\begin{equation*}
    Q = \ket{\pm}_1 \bra{\pm}_1 + \ket{\pm}_3 \bra{\pm}_3,
\end{equation*}
and 
\begin{multline*}
    \frac{1}{\pm t_2 - H_0} P = \frac{1}{\pm 2t_2} \ket{\mp}_1 \bra{\mp}_1 + \frac{1}{\pm t_2 \mp t_1} \ket{\pm}_2 \bra{\pm}_2 \\ + \frac{1}{\pm t_2 \pm t_1} \ket{\mp}_2 \bra{\mp}_2 + \frac{1}{\pm 2t_2} \ket{\mp}_3 \bra{\mp}_3.
\end{multline*}
Now we can compute
\begin{multline*}
    t'_2 = \bra{\pm}_1 H^{(2)}_\text{eff} \ket{\pm}_3 \\ = \bra{\pm}_1 H_1 P \frac{1}{\pm t_2-H_0} H_1 \ket{\pm}_3 = \pm \frac{t_0^2 t_1}{2(t_2^2-t_1^2)}.
\end{multline*}
To leading order in $\rho$ this can be expanded as
\begin{equation*}
    t'_2 = \pm \frac{t_0^2 t_1}{2 t_2 (t_2^2-t_1^2)} t_2 \approx \pm \frac{t_0^2 t_1}{2 t_2^3} t_2 = \pm \frac{\rho^{2p+3q}}{2} t_2,
\end{equation*}
from which we can read off $c_2(p,q) = \rho^{2p+3q}/2$. Finally, we can compute the ratio
\begin{multline*}
    |t'_1/t'_2| = \frac{t_0^3 t_1^2}{2(t_1^2-t_2^2)^2} \frac{2(t_2^2-t_1^2)}{t_0^2 t_1} \\ = \frac{t_0 t_1}{t_2^2-t_1^2} \approx \frac{t_0 t_1}{t_2^2} = \rho^{p+2q},
\end{multline*}
up to leading order in $\rho$.

\subsubsection{Atoms}

\paragraph{Computation of $t'_0$}
Consider the top chain in Fig.~\ref{fig:atom_molec_chains}(c), for which the Hamiltonian $H=H_0+H_1$ reads
\begin{align*}
    H_0 =& t_1 \brac{ \ket{2}\bra{3} +  \ket{6}\bra{7} + \ket{8}\bra{9} + \ket{12}\bra{13} } \\ & +t_2 \ket{4}\bra{5} +t_2 \ket{10}\bra{11} + H.c., \\
    H_1 =& t_0 \big( \ket{1}\bra{2} + \ket{3}\bra{4} + \ket{5}\bra{6} + \ket{7}\bra{8} \\ &+ \ket{9}\bra{10} + \ket{11}\bra{12} + \ket{13}\bra{14}\big) + H.c.,
\end{align*}
and the fourteen eigenstates of $H_0$ read
\begin{align*}
    \ket{\psi}_1 &=  \ket{1}, & \qquad E_0 &= 0,\\
    \ket{\pm}_2 &= \brac{\ket{2} \pm \ket{3}} / \sqrt{2}, & \qquad E_0 &= \pm t_1, \\ 
    \ket{\pm}_3 &=  \brac{\ket{4} \pm \ket{5}}/\sqrt{2}, & \qquad E_0 &= \pm t_2,\\
    \ket{\pm}_4 &= \brac{\ket{6} \pm \ket{7}} / \sqrt{2}, & \qquad E_0 &= \pm t_1, \\ 
    \ket{\pm}_5 &=  \brac{\ket{8} \pm \ket{9}}/\sqrt{2}, & \qquad E_0 &= \pm t_1,\\
    \ket{\pm}_6 &= \brac{\ket{10} \pm \ket{11}} / \sqrt{2}, & \qquad E_0 &= \pm t_2, \\ 
    \ket{\pm}_7 &=  \brac{\ket{12} \pm \ket{13}}/\sqrt{2}, & \qquad E_0 &= \pm t_1,\\
    \ket{\psi}_8 &=  \ket{14}, & \qquad E_0 &= 0,
\end{align*}
from which we can read off
\begin{equation*}
    Q = \ket{\psi}_1 \bra{\psi}_1 + \ket{\psi}_8 \bra{\psi}_8,
\end{equation*}
and 
\begin{multline*}
    \frac{1}{- H_0} P = \sum_{i=2,4,5,7} \frac{1}{\mp t_1} \ket{\pm}_i \bra{\pm}_i + \frac{1}{\pm t_1} \ket{\mp}_i \bra{\mp}_i \\ + \sum_{j=3,6} \frac{1}{\mp t_2} \ket{\pm}_j \bra{\pm}_j + \frac{1}{\pm t_2} \ket{\mp}_j \bra{\mp}_j 
\end{multline*}
Now we can compute
\begin{multline*}
    t'_0 = \bra{\psi}_1 H^{(7)}_\text{eff} \ket{\psi}_8 \\ = \bra{\psi}_1 H_1 \brac{P \frac{1}{-H_0}H_1}^{6} \ket{\psi}_8 = \frac{t_0^7}{t_1^4 t_2^2},
\end{multline*}
from which we can read off $c_0(p,q) = \frac{t_0^6}{t_1^4 t_2^2} = \rho^{6p+2q}$.

\paragraph{Computation of $t'_1$}
Consider the middle chain in Fig.~\ref{fig:atom_molec_chains}(c), for which the Hamiltonian $H=H_0+H_1$ reads
\begin{align*}
    H_0 =& t_1 \brac{ \ket{2}\bra{3} +  \ket{6}\bra{7} + \ket{10}\bra{11} } \\ & +t_2 \ket{4}\bra{5} +t_2 \ket{8}\bra{9} + H.c., \\
    H_1 =& t_0 \big( \ket{1}\bra{2} + \ket{3}\bra{4} + \ket{5}\bra{6} + \ket{7}\bra{8} \\ &+ \ket{9}\bra{10} + \ket{11}\bra{12} \big) + H.c.,
\end{align*}
and the twelve eigenstates of $H_0$ read
\begin{align*}
    \ket{\psi}_1 &=  \ket{1}, & \qquad E_0 &= 0,\\
    \ket{\pm}_2 &= \brac{\ket{2} \pm \ket{3}} / \sqrt{2}, & \qquad E_0 &= \pm t_1, \\ 
    \ket{\pm}_3 &=  \brac{\ket{4} \pm \ket{5}}/\sqrt{2}, & \qquad E_0 &= \pm t_2,\\
    \ket{\pm}_4 &= \brac{\ket{6} \pm \ket{7}} / \sqrt{2}, & \qquad E_0 &= \pm t_1, \\ 
    \ket{\pm}_5 &=  \brac{\ket{8} \pm \ket{9}}/\sqrt{2}, & \qquad E_0 &= \pm t_2,\\
    \ket{\pm}_6 &= \brac{\ket{10} \pm \ket{11}} / \sqrt{2}, & \qquad E_0 &= \pm t_1, \\ 
    \ket{\psi}_7 &=  \ket{12}, & \qquad E_0 &= 0,
\end{align*}
from which we can read off
\begin{equation*}
    Q = \ket{\psi}_1 \bra{\psi}_1 + \ket{\psi}_7 \bra{\psi}_7,
\end{equation*}
and 
\begin{multline*}
    \frac{1}{- H_0} P = \sum_{i=2,4,6} \frac{1}{\mp t_1} \ket{\pm}_i \bra{\pm}_i + \frac{1}{\pm t_1} \ket{\mp}_i \bra{\mp}_i \\ + \sum_{j=3,5} \frac{1}{\mp t_2} \ket{\pm}_j \bra{\pm}_j + \frac{1}{\pm t_2} \ket{\mp}_j \bra{\mp}_j 
\end{multline*}
Now we can compute
\begin{multline*}
    t'_1 = \bra{\psi}_1 H^{(6)}_\text{eff} \ket{\psi}_7 \\ = \bra{\psi}_1 H_1 \brac{P \frac{1}{-H_0}H_1}^{5} \ket{\psi}_7 = -\frac{t_0^6}{t_1^3 t_2^2},
\end{multline*}
from which we can read off $c_1(p,q) = \frac{t_0^6}{t_1^4 t_2^2} = \rho^{6p+2q}$. Finally, we can compute the ratio
\begin{equation*}
    |t'_0/t'_1| = \frac{t_0^7}{t_1^4 t_2^2} \frac{t_1^3 t_2^2}{t_0^6} = \frac{t_0}{t_1} = \rho^p.
\end{equation*}

\paragraph{Computation of $t'_2$}
Consider the bottom chain in Fig.~\ref{fig:atom_molec_chains}(c), for which the Hamiltonian $H=H_0+H_1$ reads
\begin{align*}
    H_0 =& t_1 \ket{2}\bra{3} + t_1 \ket{6}\bra{7}  \\ & +t_2 \ket{4}\bra{5} + H.c., \\
    H_1 =& t_0 \big( \ket{1}\bra{2} + \ket{3}\bra{4} + \ket{5}\bra{6} + \ket{7}\bra{8}  \big) + H.c.,
\end{align*}
and the eight eigenstates of $H_0$ read
\begin{align*}
    \ket{\psi}_1 &=  \ket{1}, & \qquad E_0 &= 0,\\
    \ket{\pm}_2 &= \brac{\ket{2} \pm \ket{3}} / \sqrt{2}, & \qquad E_0 &= \pm t_1, \\ 
    \ket{\pm}_3 &=  \brac{\ket{4} \pm \ket{5}}/\sqrt{2}, & \qquad E_0 &= \pm t_2,\\
    \ket{\pm}_4 &= \brac{\ket{6} \pm \ket{7}} / \sqrt{2}, & \qquad E_0 &= \pm t_1, \\ 
    \ket{\psi}_5 &=  \ket{8}, & \qquad E_0 &= 0,
\end{align*}
from which we can read off
\begin{equation*}
    Q = \ket{\psi}_1 \bra{\psi}_1 + \ket{\psi}_5 \bra{\psi}_5,
\end{equation*}
and 
\begin{multline*}
    \frac{1}{- H_0} P = \sum_{i=2,4} \frac{1}{\mp t_1} \ket{\pm}_i \bra{\pm}_i + \frac{1}{\pm t_1} \ket{\mp}_i \bra{\mp}_i \\ + \frac{1}{\mp t_2} \ket{\pm}_3 \bra{\pm}_3 + \frac{1}{\pm t_2} \ket{\mp}_3 \bra{\mp}_3 
\end{multline*}
Now we can compute
\begin{multline*}
    t'_2 = \bra{\psi}_1 H^{(4)}_\text{eff} \ket{\psi}_5 \\ = \bra{\psi}_1 H_1 \brac{P \frac{1}{-H_0}H_1}^{3} \ket{\psi}_5 = -\frac{t_0^4}{t_1^2 t_2},
\end{multline*}
from which we can read off $c_2(p,q) = \frac{t_0^4}{t_1^2 t_2^2} = \rho^{4p+2q}$. Finally, we can compute the ratio
\begin{equation*}
    |t'_1/t'_2| = \frac{t_0^6}{t_1^3 t_2^2} \frac{t_1^2 t_2}{t_0^4} = \frac{t_0^2}{t_1 t_2} = \rho^{2p+q}.
\end{equation*}

\subsection{On-Site Model}\label{App:BWPT_onsite}

The results of the perturbative calculations for the renormalized couplings $t'_i$ are summarized in Table~\ref{tab:tribo_renorm_couplings_HTC}. The construction of Eq.~(\ref{eq:tribo_renorm_result_onsite}) from Table~\ref{tab:tribo_onsite_to_hopping} is entirely analogous to the HTC case. Since the values $a_i,b_i$ in the last two columns are not represented as powers of $\rho$, we use the mathematical identity $a = \rho^{\log a /\log \rho}$ for any real number $a>0$. Using this trick, the values $a_i,b_i$ can be converted to the exponents $p_i = \log a_i / \log \rho$ and $q_i = \log b_i / \log \rho$ in Eq.~(\ref{eq:tribo_renorm_result_onsite}). 


\begin{figure*}[tb]
    \includegraphics[width = \linewidth]{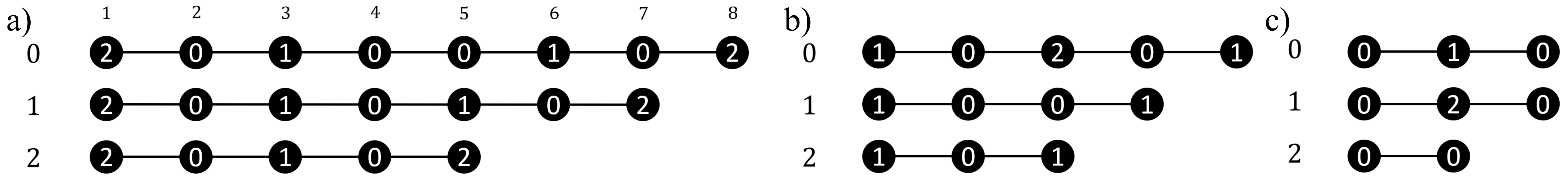}
    \caption{The different chains occurring in the OTC. The black dots denote the lattice sites, the black lines a bond $t$ and the number in the black dot denotes $\epsilon_i$, the on-site potential . a) the type-2 atomic chains, b) type-1 atomic chains and c) type-0 atomic chains. The vertical numbers $0,1,2$ denote the letter to which that chain renormalizes, the horizontal numbers $1,2,\dots$ denote the lattice site labeling in the perturbative calculations in Section~\ref{App:BWPT_onsite}.}
    \label{fig:atom_molec_chains_onsite}
\end{figure*}

The remainder of this section is devoted to computing the $t'_i$ values in Table~\ref{tab:tribo_onsite_to_hopping}. All calculations are done for the OTC $H^o_N$, for arbitrary real $c_1$ and $c_2$ that satisfy $|c_1| \gg 1, |c_2| \gg 1$ and $|c_2-c_1| \gg 1$. The perturbative calculations will be easier than for the hopping model, since the unperturbed OTC is diagonal.

\subsubsection{Type-2 Atoms}

\paragraph{Computation of $t'_0$}
Consider the top chain in Fig.~\ref{fig:atom_molec_chains_onsite}(a), for which the Hamiltonian $H=H_0+H_1$ reads
\begin{align*}
    &H_0 = \epsilon_2 \ket{1}\bra{1} + \epsilon_0 \ket{2}\bra{2} + \epsilon_1 \ket{3}\bra{3}  + \epsilon_0 \ket{4}\bra{4} \\ &+ \epsilon_0 \ket{5}\bra{5} + \epsilon_1 \ket{6}\bra{6} + \epsilon_0 \ket{7}\bra{7} + \epsilon_2 \ket{8}\bra{8}  + H.c., \\
    &H_1 = \sum^7_{i=1} t \ket{i}\bra{i+1} + H.c.,
\end{align*}
and the eight eigenstates of $H_0$ read $H_0 \ket{i} = E_0 \ket{i}$, from which we can read off
\begin{equation*}
    Q = \ket{1} \bra{1} + \ket{8} \bra{8},
\end{equation*}
and the operator
\begin{multline*}
    \frac{1}{\epsilon_2 - H_0} P = \\ \frac{1}{\epsilon_2 - \epsilon_0}\ket{2}\bra{2} + \frac{1}{\epsilon_2 - \epsilon_1}\ket{3}\bra{3} + \frac{1}{\epsilon_2 - \epsilon_0}\ket{4}\bra{4}  \\ + \frac{1}{\epsilon_2 - \epsilon_0}\ket{5}\bra{5}  + \frac{1}{\epsilon_2 - \epsilon_1}\ket{6}\bra{6} + \frac{1}{\epsilon_2 - \epsilon_0}\ket{7}\bra{7},
\end{multline*}
which is also diagonal. Now we can compute 
\begin{multline*}
    t'_0 = \bra{1} H^{(7)}_\text{eff} \ket{8} = \bra{1} H_1 \brac{P \frac{1}{\epsilon_2-H_0}H_1}^{6} \ket{8} \\ =   \frac{t^7}{(\epsilon_2 - \epsilon_0)^4 (\epsilon_2-\epsilon_1)^2} = \frac{t}{c_2^4 (c_2-c_1)^2}.
\end{multline*}

\paragraph{Computation of $t'_1$}
Consider the middle chain in Fig.~\ref{fig:atom_molec_chains_onsite}(a), for which the Hamiltonian $H=H_0+H_1$ reads
\begin{align*}
    H_0 =& \, \epsilon_2 \ket{1}\bra{1} + \epsilon_0 \ket{2}\bra{2} + \epsilon_1 \ket{3}\bra{3}  + \epsilon_0 \ket{4}\bra{4}  \\ &+ \epsilon_1 \ket{5}\bra{5} + \epsilon_0 \ket{6}\bra{6} + \epsilon_2 \ket{7}\bra{7}  + H.c., \\
    H_1 =& \sum^6_{i=1} t \ket{i}\bra{i+1} + H.c.,
\end{align*}
and the seven eigenstates of $H_0$ read $H_0 \ket{i} = E_0 \ket{i}$, from which we can read off
\begin{equation*}
    Q = \ket{1} \bra{1} + \ket{7} \bra{7},
\end{equation*}
and the operator
\begin{multline*}
    \frac{1}{\epsilon_2 - H_0} P = \frac{1}{\epsilon_2 - \epsilon_0}\ket{2}\bra{2} + \frac{1}{\epsilon_2 - \epsilon_1}\ket{3}\bra{3} \\ + \frac{1}{\epsilon_2 - \epsilon_0}\ket{4}\bra{4} + \frac{1}{\epsilon_2 - \epsilon_1}\ket{5}\bra{5} + \frac{1}{\epsilon_2 - \epsilon_0}\ket{6}\bra{6},
\end{multline*}
which is also diagonal. Now we can compute 
\begin{multline*}
    t'_1 = \bra{1} H^{(6)}_\text{eff} \ket{7} = \bra{1} H_1 \brac{P \frac{1}{\epsilon_2-H_0}H_1}^{5} \ket{7} \\ =   \frac{t^6}{(\epsilon_2 - \epsilon_0)^3 (\epsilon_2-\epsilon_1)^2} = \frac{t}{c_2^3 (c_2-c_1)^2}.
\end{multline*}
Finally, we can compute the ratio
\begin{equation*}
    |t'_0/t'_1| = \left| \frac{t}{c_2^4 (c_2-c_1)^2} \frac{c_2^3 (c_2-c_1)^2}{t} \right| = \frac{1}{|c_2|}.
\end{equation*}

\paragraph{Computation of $t'_2$}
Consider the bottom chain in Fig.~\ref{fig:atom_molec_chains_onsite}(a), for which the Hamiltonian $H=H_0+H_1$ reads
\begin{align*}
    H_0 =& \, \epsilon_2 \ket{1}\bra{1} + \epsilon_0 \ket{2}\bra{2} + \epsilon_1 \ket{3}\bra{3}  + \epsilon_0 \ket{4}\bra{4}  \\ &+ \epsilon_2 \ket{5}\bra{5}  + H.c., \\
    H_1 =& \sum^4_{i=1} t \ket{i}\bra{i+1} + H.c.,
\end{align*}
and the five eigenstates of $H_0$ read $H_0 \ket{i} = E_0 \ket{i}$, from which we can read off
\begin{equation*}
    Q = \ket{1} \bra{1} + \ket{5} \bra{5},
\end{equation*}
and the operator
\begin{multline*}
    \frac{1}{\epsilon_2 - H_0} P = \frac{1}{\epsilon_2 - \epsilon_0}\ket{2}\bra{2} \\ + \frac{1}{\epsilon_2 - \epsilon_1}\ket{3}\bra{3} + \frac{1}{\epsilon_2 - \epsilon_0}\ket{4}\bra{4},
\end{multline*}
which is also diagonal. Now we can compute 
\begin{multline*}
    t'_2 = \bra{1} H^{(4)}_\text{eff} \ket{5} = \bra{1} H_1 \brac{P \frac{1}{\epsilon_2-H_0}H_1}^{3} \ket{5} \\ =   \frac{t^4}{(\epsilon_2 - \epsilon_0)^2 (\epsilon_2-\epsilon_1)} = \frac{t}{c_2 (c_2-c_1)}.
\end{multline*}
Finally, we can compute the ratio
\begin{equation*}
    |t'_1/t'_2| = \left|  \frac{t}{c_2^3 (c_2-c_1)^2} \frac{c_2 (c_2-c_1)}{t} \right| = \frac{1}{|c_2 (c_2-c_1)|}.
\end{equation*}

\begin{table*}[htb]
\caption{\label{tab:tribo_onsite_to_hopping}Effective hopping parameters for the OTC.}
\centering
\begin{tabular}{||c | c c c | c c ||} 
 \hline
 $E_0$  & $t'_0$ & $t'_1$ & $t'_2$  & $a_i = |t'_0/t'_1|$ & $b_i = |t'_1/t'_2|$  \\ [0.5ex] 
 \hline\hline
 $\epsilon_0$ & $t^2/(\epsilon_0-\epsilon_1)$ & $t^2/(\epsilon_0-\epsilon_2)$ & $t$ & $|c_2/c_1|$ & $1/|c_2|$ \\ 
 \hline
 $\epsilon_1$ & $t^4/(\epsilon_1-\epsilon_0)^2(\epsilon_1-\epsilon_2)$ & $t^3/(\epsilon_1-\epsilon_0)^2$ & $t^2/(\epsilon_1-\epsilon_0)$ & $1/|c_2-c_1|$ & $1/|c_1|$    \\
 \hline
 $\epsilon_2$ & $t^7/(\epsilon_2-\epsilon_0)^4(\epsilon_2-\epsilon_1)^2$ & $t^6/(\epsilon_2-\epsilon_0)^3 (\epsilon_2-\epsilon_1)^2$ & $t^4/(\epsilon_2-\epsilon_0)^2 (\epsilon_2-\epsilon_1)$ & $1/|c_2|$ & $1/|c_2(c_2-c_1)|$ \\
 \hline
\end{tabular}
\end{table*}

\subsubsection{Type-1 Atoms}

\paragraph{Computation of $t'_0$}
Consider the top chain in Fig.~\ref{fig:atom_molec_chains_onsite}(b), for which the Hamiltonian $H=H_0+H_1$ reads
\begin{align*}
    H_0 =& \, \epsilon_1 \ket{1}\bra{1} + \epsilon_0 \ket{2}\bra{2} + \epsilon_2 \ket{3}\bra{3}  + \epsilon_0 \ket{4}\bra{4}  \\ &+ \epsilon_1 \ket{5}\bra{5}  + H.c., \\
    H_1 =& \sum^4_{i=1} t \ket{i}\bra{i+1} + H.c.,
\end{align*}
and the five eigenstates of $H_0$ read $H_0 \ket{i} = E_0 \ket{i}$, from which we can read off
\begin{equation*}
    Q = \ket{1} \bra{1} + \ket{5} \bra{5},
\end{equation*}
and the operator
\begin{multline*}
    \frac{1}{\epsilon_1 - H_0} P = \frac{1}{\epsilon_1 - \epsilon_0}\ket{2}\bra{2} \\ + \frac{1}{\epsilon_1 - \epsilon_2}\ket{3}\bra{3} + \frac{1}{\epsilon_1 - \epsilon_0}\ket{4}\bra{4},
\end{multline*}
which is also diagonal. Now we can compute 
\begin{multline*}
    t'_0 = \bra{1} H^{(4)}_\text{eff} \ket{5} = \bra{1} H_1 \brac{P \frac{1}{\epsilon_1-H_0}H_1}^{3} \ket{5} \\ =   \frac{t^4}{(\epsilon_1 - \epsilon_0)^2 (\epsilon_1-\epsilon_2)} = -\frac{t}{c_1^2 (c_2-c_1)}.
\end{multline*}

\paragraph{Computation of $t'_1$}
Consider the middle chain in Fig.~\ref{fig:atom_molec_chains_onsite}(b), for which the Hamiltonian $H=H_0+H_1$ reads
\begin{align*}
    H_0 =& \, \epsilon_1 \ket{1}\bra{1} + \epsilon_0 \ket{2}\bra{2} + \epsilon_0 \ket{3}\bra{3}  + \epsilon_1 \ket{4}\bra{4}  + H.c., \\
    H_1 =& \sum^3_{i=1} t \ket{i}\bra{i+1} + H.c.,
\end{align*}
and the four eigenstates of $H_0$ read $H_0 \ket{i} = E_0 \ket{i}$, from which we can read off
\begin{equation*}
    Q = \ket{1} \bra{1} + \ket{4} \bra{4},
\end{equation*}
and the operator
\begin{multline*}
    \frac{1}{\epsilon_1 - H_0} P = \frac{1}{\epsilon_1 - \epsilon_0}\ket{2}\bra{2}  + \frac{1}{\epsilon_1 - \epsilon_0}\ket{3}\bra{3} ,
\end{multline*}
which is also diagonal. Now we can compute 
\begin{multline*}
    t'_1 = \bra{1} H^{(3)}_\text{eff} \ket{4} = \bra{1} H_1 \brac{P \frac{1}{\epsilon_1-H_0}H_1}^{2} \ket{4} \\ =   \frac{t^3}{(\epsilon_1 - \epsilon_0)^2} = \frac{t}{c_1^2}.
\end{multline*}
Finally, we can compute the ratio
\begin{equation*}
    |t'_0/t'_1| = \left| \frac{t}{c_1^2 (c_2-c_1)} \frac{c_1^2 }{t} \right| = \frac{1}{|c_2-c_1|}.
\end{equation*}

\paragraph{Computation of $t'_2$}
Consider the bottom chain in Fig.~\ref{fig:atom_molec_chains_onsite}(b), for which the Hamiltonian $H=H_0+H_1$ reads
\begin{align*}
    H_0 =& \, \epsilon_1 \ket{1}\bra{1} + \epsilon_0 \ket{2}\bra{2} + \epsilon_1 \ket{3}\bra{3} + H.c., \\
    H_1 =&  t \ket{1}\bra{2} + t \ket{2}\bra{3} + H.c.,
\end{align*}
and the three eigenstates of $H_0$ read $H_0 \ket{i} = E_0 \ket{i}$, from which we can read off
\begin{equation*}
    Q = \ket{1} \bra{1} + \ket{3} \bra{3},
\end{equation*}
and the operator
\begin{equation*}
    \frac{1}{\epsilon_1 - H_0} P = \frac{1}{\epsilon_1 - \epsilon_0}\ket{2}\bra{2} ,
\end{equation*}
which is also diagonal. Now we can compute 
\begin{multline*}
    t'_2 = \bra{1} H^{(2)}_\text{eff} \ket{3} = \bra{1} H_1 P \frac{1}{\epsilon_1-H_0}H_1 \ket{3} \\ =   \frac{t^2}{\epsilon_1 - \epsilon_0} = \frac{t}{c_1}.
\end{multline*}
Finally, we can compute the ratio
\begin{equation*}
    |t'_1/t'_2| = \left| \frac{t}{c_1^2 } \frac{c_1}{t} \right| = \frac{1}{|c_1|}.
\end{equation*}

\subsubsection{Type-0 Atoms}

\paragraph{Computation of $t'_0$}
Consider the top chain in Fig.~\ref{fig:atom_molec_chains_onsite}(c), for which the Hamiltonian $H=H_0+H_1$ reads
\begin{align*}
    H_0 =& \, \epsilon_0 \ket{1}\bra{1} + \epsilon_1 \ket{2}\bra{2} + \epsilon_0 \ket{3}\bra{3} + H.c., \\
    H_1 =&  t \ket{1}\bra{2} + t \ket{2}\bra{3} + H.c.,
\end{align*}
and the three eigenstates of $H_0$ read $H_0 \ket{i} = E_0 \ket{i}$, from which we can read off
\begin{equation*}
    Q = \ket{1} \bra{1} + \ket{3} \bra{3},
\end{equation*}
and the operator
\begin{equation*}
    \frac{1}{\epsilon_0 - H_0} P = \frac{1}{\epsilon_0 - \epsilon_1}\ket{2}\bra{2} ,
\end{equation*}
which is also diagonal. Now we can compute 
\begin{multline*}
    t'_0 = \bra{1} H^{(2)}_\text{eff} \ket{3} = \bra{1} H_1 P \frac{1}{\epsilon_0-H_0}H_1 \ket{3} \\ =   \frac{t^2}{\epsilon_0 - \epsilon_1} = -\frac{t}{c_1}.
\end{multline*}

\paragraph{Computation of $t'_1$}
Consider the middle chain in Fig.~\ref{fig:atom_molec_chains_onsite}(c), for which the Hamiltonian $H=H_0+H_1$ reads
\begin{align*}
    H_0 =& \, \epsilon_0 \ket{1}\bra{1} + \epsilon_2 \ket{2}\bra{2} + \epsilon_0 \ket{3}\bra{3} + H.c., \\
    H_1 =&  t \ket{1}\bra{2} + t \ket{2}\bra{3} + H.c.,
\end{align*}
and the three eigenstates of $H_0$ read $H_0 \ket{i} = E_0 \ket{i}$, from which we can read off
\begin{equation*}
    Q = \ket{1} \bra{1} + \ket{3} \bra{3},
\end{equation*}
and the operator
\begin{equation*}
    \frac{1}{\epsilon_0 - H_0} P = \frac{1}{\epsilon_0 - \epsilon_2}\ket{2}\bra{2} ,
\end{equation*}
which is also diagonal. Now we can compute 
\begin{multline*}
    t'_1 = \bra{1} H^{(2)}_\text{eff} \ket{3} = \bra{1} H_1 P \frac{1}{\epsilon_0-H_0}H_1 \ket{3} \\ =   \frac{t^2}{\epsilon_0 - \epsilon_2} = -\frac{t}{c_2}.
\end{multline*}
Finally, we can compute the ratio $|t'_0/t'_1| = \left| c_2/c_1 \right|$.

\paragraph{Computation of $t'_2$}
Consider the bottom chain in Fig.~\ref{fig:atom_molec_chains_onsite}(c), for which the Hamiltonian $H=H_0+H_1$ reads
\begin{align*}
    H_0 =& \, \epsilon_0 \ket{1}\bra{1} + \epsilon_0 \ket{2}\bra{2}  + H.c., \\
    H_1 =&  t \ket{1}\bra{2} + H.c.,
\end{align*}
and the two eigenstates of $H_0$ read $H_0 \ket{i} = \epsilon_0 \ket{i}$, from which we can read off
\begin{equation*}
    Q = \ket{1} \bra{1} + \ket{2} \bra{2} = \operatorname{Id}, \qquad \frac{1}{\epsilon_0 - H_0} P = 0,
\end{equation*}
which are not needed anyway. Now we can compute 
\begin{equation*}
    t'_2 = \bra{1} H^{(1)}_\text{eff} \ket{2} = \bra{1} H_1  \ket{2} = t,
\end{equation*}
and the ratio $|t'_1/t'_2| =  1/|c_2| $.

\section{Proof of Equivalence of HTC and OTC}\label{app:equivalence_TC_OTC}

This section is devoted to proving Eq.~(\ref{eq:tribo_equiv_goal}). The dynamical system on the values $p,q$ generated by Eq.~(\ref{eq:tribo_renorm_result}) can be represented by the map
\begin{equation}\label{eq:pq_DS_original}
    f:\R^2 \to \R^{10}, \qquad (p,q) \mapsto 
\begin{Bmatrix}
(p+q,p+2q) \\
(q,p) \\
(p,2p+q) \\
(q,p) \\
(p+q,p+2q) 
\end{Bmatrix}.
\end{equation}
We can modify Eq.~(\ref{eq:pq_DS_original}) to the following function, that yields the same, or smaller values:
\begin{equation}\label{eq:pq_DS_edit1}
    \Tilde{f}:\R^2 \to \R^{10}, \qquad (p,q) \mapsto 
\begin{Bmatrix}
(2 \Tilde{p},2 \Tilde{p}) \\
( \Tilde{p}, \Tilde{p}) \\
( \Tilde{p}, \Tilde{p}) \\
( \Tilde{p}, \Tilde{p}) \\
(2 \Tilde{p}, 2 \Tilde{p}) 
\end{Bmatrix}, 
\end{equation}
where $\Tilde{p} := \min \{ p,q \}$. Note that Eq.~(\ref{eq:pq_DS_edit1}) is symmetric with respect to the permutation $p \xleftrightarrow{} q$ in its arguments, as well as in each of the five tuples of the output, which allows us to write as a function of only one variable:
\begin{equation}\label{eq:pq_DS_edit2}
    g : \R \to \R^{5}, \qquad \Tilde{p} \mapsto \{ 2\Tilde{p},\Tilde{p},\Tilde{p},\Tilde{p},2\Tilde{p} \}.
\end{equation}
Starting with any $p,q,\Tilde{p} \in \R$, note that after $N$ applications of $f $ to $p,q$ and $g$ to $\Tilde{p}$, it is true that $p_i^{(N)},q_i^{(N)} \geq \Tilde{p}^{(N)}_i$, for each $i$. Note that after $N$ applications of $g$ to some initial value $\Tilde{p}$, we have a set $I^g_N$ of $5^N$ different real values. Similarly, define $J^g_N := \{ x \in I^g_N \mid x \leq N \} $, we can now rigorously show $\lim_{N \to \infty} |J^g_N|/|I^g_N| = 0$, which implies that Eq.~(\ref{eq:tribo_equiv_goal}) holds. Note that the map $g$ leaves $3/5$ of its output constant and multiplies $2/5$ with a factor of two. For $x \in J^g_N$ to hold, it can be multiplied by a factor of two at most $\lfloor \log_2 N \rfloor$ times. By this argument, $|J^g_N|$ can exactly be computed by counting the possible ways that $g$ can be successively applied to a single initial value $\Tilde{p}_0>0$. First let $\Tilde{n} = \lceil \log_2 \Tilde{p}_0 \rceil$, such that $2^{\Tilde{n}} \geq \Tilde{p}_0$. Then we have the exact combinatorial argument
\begin{multline*}
    |J^g_N|/|I^g_N| = \sum^{\lfloor \log_2 N \rfloor - \Tilde{n}}_{n=0} \binom{N}{n} (3/5)^{N - n } (2/5)^{n} \\ 
    < \sum^{\lfloor \log_2 N \rfloor}_{n=0} \binom{N}{n} (3/5)^{N - n } (2/5)^{n}, 
\end{multline*}
which can be bounded from above as 
\begin{equation}\label{eq:pq_intermediate1}
    \sum^{\lfloor \log_2 N \rfloor}_{n=0} \binom{N}{n} (3/5)^{N - n } (2/5)^{n} \leq \binom{N}{\log_2 N} (3/5)^N  \log_2 N.
\end{equation}
Now we need to approximate $\binom{N}{\log_2 N}$, for which we can use Stirling's approximation $\log n! = n \log n - n + \mathcal{O}(\log n)$: 
\begin{align*}
    & \log \binom{N}{\log_2 N} = \\  
     & \log N! - \log (N-\log_2 N)!  -\log (\log_2 N)! \\
    =& N \log N - (N-\log_2 N) \log (N-\log_2 N) + \mathcal{O}(\log N)  \\
    =& N \log N - (N-\log_2 N) \brac{\log N + \log (1- \frac{\log_2 N}{N})} \\ &+ \mathcal{O}(\log N) \\
    =& \frac{\log^2 N}{\log 2} + (N-\log_2 N) \brac{  \frac{\log_2 N}{N} + \mathcal{O}\brac{\frac{\log^2_2 N}{N^2}}} \\ &+ \mathcal{O}(\log N) \\
    =& \frac{\log^2 N}{\log 2} + \mathcal{O}(\log N).
\end{align*}
Since we are interested in the large $N$ limit, it is enough to know the divergent behaviour of $\binom{N}{\log_2 N}$. We can now further approximate Eq.~(\ref{eq:pq_intermediate1}) as 
\begin{multline*}
    \binom{N}{\log_2 N} (3/5)^N  \log_2 N  \\
    = e^{\log^2 N / \log 2 + \mathcal{O}(\log n)} e^{N (\log 3  - \log 5)} e^{\log \log_2 N }  \\ 
    = e^{N (\log 3 - \log 5) + \mathcal{O}(\log^2 N)} \xrightarrow[]{N \to \infty} 0,
\end{multline*}
since $\log 3 - \log 5 < 0$. This means that 
\begin{equation*}
    0 \leq |J_N| / |I_N| \leq |J^g_N| / |I^g_N|\xrightarrow[]{N \to \infty} 0,
\end{equation*}
proving the statement in Eq.~(\ref{eq:tribo_equiv_goal}).


\section{Eigenstates on the Rauzy Fractal}\label{app:eigenstates}

\begin{figure*}[b]
    \centering
    \includegraphics[height = 0.9\textheight]{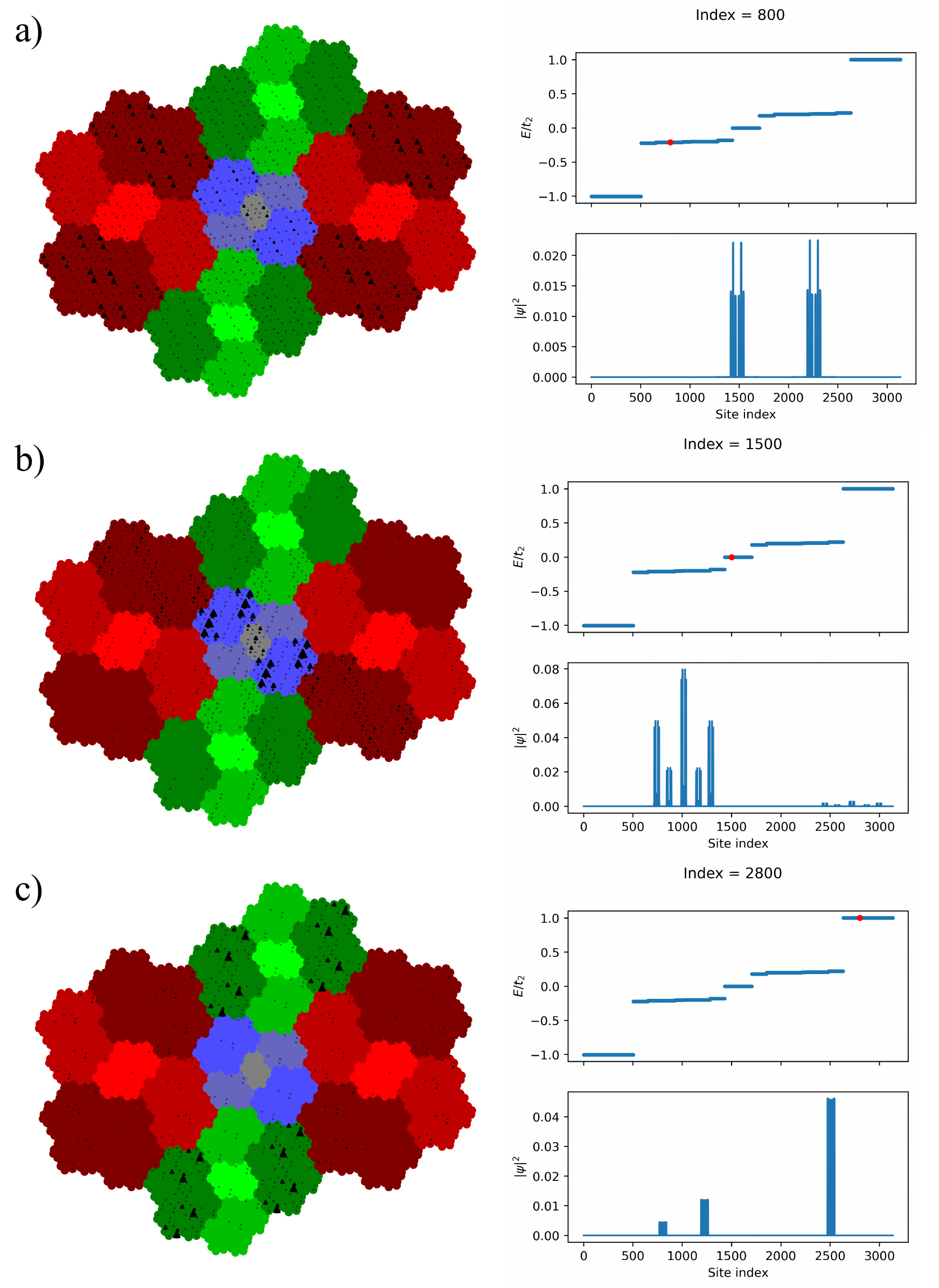}
    \caption{(Color online) Three eigenstates of the HTC $H_{13}$, plotted on the subdivided Rauzy fractal. The height of the black triangles on site $n$ are proportional to $|\psi_i(n)|^2$. On the right hand side, the eigenstate is plotted, together with the state (red dot) in the spectrum (blue). a) $i=800$, b) $i=1500$ and c) $i=2800$.} 
    \label{fig:states_rauzy_examples_HTC}
\end{figure*}

\begin{figure*}[b]
    \centering
    \includegraphics[height = 0.9\textheight]{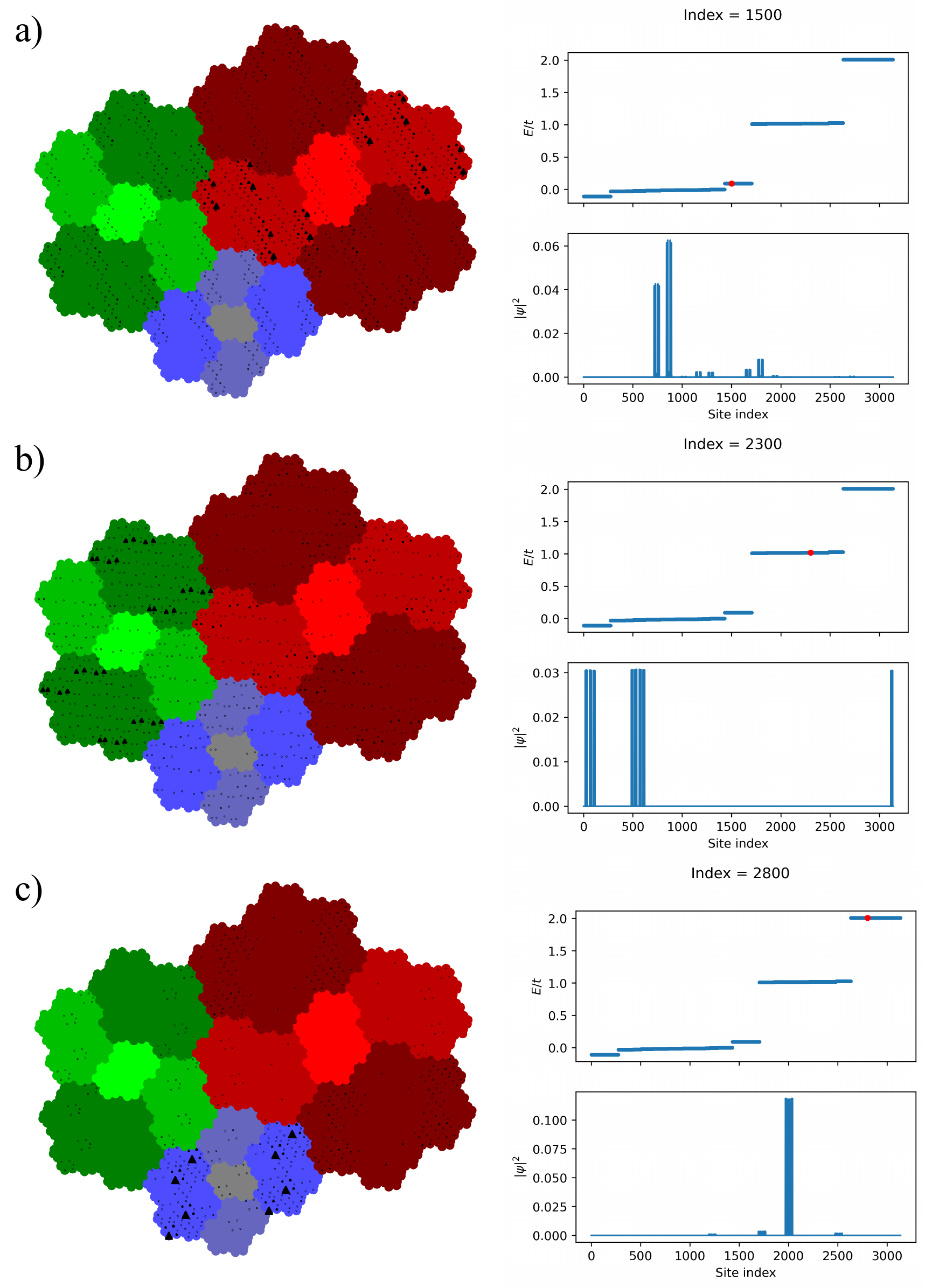}
    \caption{(Color online) Three eigenstates of the OTC $H_{13}^o$, plotted on the subdivided Rauzy fractal. The height of the black triangles on site $n$ are proportional to $|\psi_i(n)|^2$. On the right hand side, the eigenstate is plotted, together with the state (red dot) in the spectrum (blue). a) $i=1500$, b) $i=2300$ and c) $i=2800$.}  
    \label{fig:states_rauzy_examples_OTC}
\end{figure*}

Figs.~\ref{fig:states_rauzy_examples_HTC} and \ref{fig:states_rauzy_examples_OTC} contain more examples of eigenstates plotted on the Rauzy fractal, where the Rauzy fractal is subdivided again according to environments of the local structures. These figures give more evidence for the observation that the eigenstates localize on the Rauzy fractal in regions that correspond to the branch of the spectrum of the eigenstate. For example, in Fig.~\ref{fig:states_rauzy_examples_OTC}(a)/(b)/(c), the state belongs to the bottom/middle/top branch of the spectrum, hence localizes on the red/green/blue area of the Rauzy fractal. 

\end{document}